\documentclass[proof]{pasj01}

\usepackage{graphicx}
\usepackage{natbib}
\usepackage{multirow}
\usepackage{listings}

\begin{document}
\Received{}
\Accepted{}
\Published{}

\title{Atmospheric gas dynamics in the Perseus cluster observed with Hitomi}




\author{Hitomi Collaboration\thanks{The corresponding authors are
Yuto \textsc{Ichinohe},
Shutaro \textsc{Ueda},
Ryuichi \textsc{Fujimoto},
Shota \textsc{Inoue},
Caroline \textsc{Kilbourne},
Tetsu \textsc{Kitayama},
Maxim \textsc{Markevitch},
Brian \textsc{McNamara},
Naomi \textsc{Ota},
Scott \textsc{Porter},
Takayuki \textsc{Tamura},
Keigo \textsc{Tanaka}
and
Norbert \textsc{Werner}}
,
Felix \textsc{Aharonian}\altaffilmark{1,2,3},
Hiroki \textsc{Akamatsu}\altaffilmark{4},
Fumie \textsc{Akimoto}\altaffilmark{5},
Steven W. \textsc{Allen}\altaffilmark{6,7,8},
Lorella \textsc{Angelini}\altaffilmark{9},
Marc \textsc{Audard}\altaffilmark{10},
Hisamitsu \textsc{Awaki}\altaffilmark{11},
Magnus \textsc{Axelsson}\altaffilmark{12},
Aya \textsc{Bamba}\altaffilmark{13,14},
Marshall W. \textsc{Bautz}\altaffilmark{15}
% ...
%
%
}

\altaffiltext{1}{Dublin Institute for Advanced Studies, 31 Fitzwilliam Place, Dublin 2, Ireland}
\altaffiltext{2}{Max-Planck-Institut f{\"u}r Kernphysik, P.O. Box 103980, 69029 Heidelberg, Germany}
\altaffiltext{3}{Gran Sasso Science Institute, viale Francesco Crispi, 7 67100 L'Aquila (AQ), Italy}
\altaffiltext{4}{SRON Netherlands Institute for Space Research, Sorbonnelaan 2, 3584 CA Utrecht, The Netherlands}
\altaffiltext{5}{Institute for Space-Earth Environmental Research, Nagoya University, Furo-cho, Chikusa-ku, Nagoya, Aichi 464-8601}
\altaffiltext{6}{Kavli Institute for Particle Astrophysics and Cosmology, Stanford University, 452 Lomita Mall, Stanford, CA 94305, USA}
\altaffiltext{7}{Department of Physics, Stanford University, 382 Via Pueblo Mall, Stanford, CA 94305, USA}
\altaffiltext{8}{SLAC National Accelerator Laboratory, 2575 Sand Hill Road, Menlo Park, CA 94025, USA}
\altaffiltext{9}{NASA, Goddard Space Flight Center, 8800 Greenbelt Road, Greenbelt, MD 20771, USA}
\altaffiltext{10}{Department of Astronomy, University of Geneva, ch. d'\'Ecogia 16, CH-1290 Versoix, Switzerland}
\altaffiltext{11}{Department of Physics, Ehime University, Bunkyo-cho, Matsuyama, Ehime 790-8577}
\altaffiltext{12}{Department of Physics and Oskar Klein Center, Stockholm University, 106 91 Stockholm, Sweden}
\altaffiltext{13}{Department of Physics, The University of Tokyo, 7-3-1 Hongo, Bunkyo-ku, Tokyo 113-0033}
\altaffiltext{14}{Research Center for the Early Universe, School of Science, The University of Tokyo, 7-3-1 Hongo, Bunkyo-ku, Tokyo 113-0033}
\altaffiltext{15}{Kavli Institute for Astrophysics and Space Research, Massachusetts Institute of Technology, 77 Massachusetts Avenue, Cambridge, MA 02139, USA}
% ...

\email{ichinohe@tmu.ac.jp, sueda@astro.isas.jaxa.jp, fujimoto@se.kanazawa-u.ac.jp, shota@ess.sci.osaka-u.ac.jp, Caroline.A.Kilbourne@nasa.gov, kitayama@ph.sci.toho-u.ac.jp, m.markevitch@gmail.com, mcnamara@uwaterloo.ca, naomi@cc.nara-wu.ac.jp, Frederick.S.Porter@nasa.gov, tamura.takayuki@jaxa.jp, tanaka@astro.s.kanazawa-u.ac.jp, wernernorbi@gmail.com}



\maketitle

\begin{abstract}
Extending the earlier measurements reported in Hitomi collaboration (2016, Nature, 535, 117), we examine the atmospheric gas motions within the central 100~kpc of the Perseus cluster using observations obtained with the Hitomi satellite. After correcting for the point spread function of the telescope and using optically thin emission lines, we find that the line-of-sight velocity dispersion of the hot gas is remarkably low and mostly uniform. The velocity dispersion reaches maxima of approximately 200~km~s$^{-1}$ toward the central active galactic nucleus (AGN) and toward the AGN inflated north-western `ghost' bubble. Elsewhere within the observed region, the velocity dispersion appears constant around 100~km~s$^{-1}$. We also detect a velocity gradient with a 100~km~s$^{-1}$ amplitude across the cluster core, consistent with large-scale sloshing of the core gas. If the observed gas motions are isotropic, the kinetic pressure support is less than 10\% of the thermal pressure support in the cluster core. The well-resolved optically thin emission lines have Gaussian shapes, indicating that the turbulent driving scale is likely below 100~kpc, which is consistent with the size of the AGN jet inflated bubbles. We also report the first measurement of the ion temperature in the intracluster medium, which we find to be consistent with the electron temperature. In addition, we present a new measurement of the redshift to the brightest cluster galaxy NGC~1275.
\end{abstract}

\section{Introduction}
\label{sec:intro}

Clusters of galaxies are the most massive bound and virialized structures in the Universe. Their peripheries are dynamically young as clusters continue to grow through the accretion of surrounding matter. Disturbances due to subcluster mergers are found even in relaxed clusters with cool cores \citep[e.g.,][]{markevitch01,Churazov03,Clarke04,Blanton11,Ueda17}. Mergers are expected to drive shocks, bulk shear, and turbulence in the intracluster medium (ICM). Clusters with cool cores also host active galactic nuclei \citep[AGN;][]{Burns90,Sun09} which inject mechanical energy and magnetic fields into the gas of the cluster cores that drive its motions \citep[e.g.,][]{Boehringer93,Carilli94,Churazov00,McNamara00,Fabian03,Werner10}. Such AGN feedback may play a major role in preventing runaway cooling in cluster cores \citep[see][for reviews]{McNamara07,Fabian12}. Knowledge of the dynamics of the ICM will be crucial for understanding the physics of galaxy clusters such as heating and thermalization of the gas, acceleration of relativistic particles, and the level of atmospheric viscosity. It also probes the degree to which hot atmospheres are in hydrostatic balance, which has been widely assumed in cosmological studies using galaxy clusters \citep[see][for review]{Allen11}.

Bulk and turbulent motions have been difficult to measure owing to the lack of non-dispersive X-ray spectrometers with sufficient energy resolution to resolve line-of-sight (LOS) velocities. For example, a LOS bulk velocity of 500~km~s$^{-1}$ produces a Doppler shift of 11~eV for the Fe~XXV He$\alpha$ line at 6.7~keV. Most of the previous attempts using X-ray charge-coupled device (CCD) cameras, with typical energy resolutions of $\sim$150~eV, lead to upper limits or low significance ($< 3\sigma$) detections of bulk motions \citep[e.g.,][]{Dupke06,Ota07,Dupke07,Fujita08,Sato08,Sato11,Sugawara09,Nishino12,Tamura14,Ota16}; higher significance measurements were reported only in a few merging clusters \citep{Tamura11,Liu16}.

Upper limits on Doppler broadening were also obtained using the Reflection Grating Spectrometer on board XMM-Newton \citep[RGS;][]{den_Herder01} with typical values of 200--600~km~s$^{-1}$ at the 68\% confidence level \citep{Sanders10,Sanders11,bulbul12,Sanders13,Pinto15}. As the RGS is slitless, spectral lines are broadened by the spatial extent of the ICM, making it challenging to separate and spatially map the Doppler widths.

The Soft X-ray Spectrometer \citep[SXS;][]{Kelley16} on board Hitomi \citep{Takahashi16} is the first X-ray instrument in orbit capable of resolving the emission lines in extended sources and measuring their Doppler broadening and shifts. The SXS is a non-dispersive spectrometer with an energy resolution of 4.9~eV full-width at half-maximum (FWHM) at 6~keV \citep{porter16}. The SXS imaged the core of the Perseus cluster, the brightest galaxy cluster in the X-ray sky. Previous X-ray observations of this region revealed a series of faint, X-ray cavities around the AGN in the central galaxy NGC~1275 \citep{Boehringer93,McNamara96,Churazov00,Fabian00} as well as weak shocks and ripples \citep{Fabian03,Fabian06,Fabian11,Sanders07}, both suggestive of the presence of gas motions. The SXS performed four pointings in total with a field of view (FOV) of 60~kpc $\times$ 60~kpc each and a total exposure time of 320~ks as shown in figure~\ref{fig:fov} and table~\ref{tab:obs}. Early results based on two pointings toward nearly the same sky region (Obs~2 and Obs~3) were published in \citet[][hereafter H16]{hitomi16}. H16 reported that the LOS velocity dispersion in a region 30--60~kpc from the central AGN is $164 \pm 10$~km~s$^{-1}$ and the gradient in the LOS bulk velocity across the image is $150 \pm 70$~km~s$^{-1}$, where the quoted errors denote 90\% statistical uncertainties.

In this paper, we present a thorough analysis of gas motions in the Perseus cluster measured with Hitomi. Updates from H16 include; (i) the full dataset including remaining two offset pointings (Obs~1 and Obs~4) are analyzed to probe the gas motions out to 100~kpc from the central AGN; (ii) the effects of the point spread function (PSF) of the telescope with the half power diameter (HPD) of 1.2~arcmin \citep{okajima16} are taken into account in deriving the velocity maps; (iii) the absolute gas velocities are compared to a new recession velocity of NGC~1275 based on stellar absorption lines; (iv) detailed shapes of bright emission lines are examined to search for non-Gaussianity of the distribution function of the gas velocity; (v) constraints on the thermal motion of ions in the ICM are derived combining the widths of the lines originating from various elements; and (vi) revised calibration and improved estimation for the systematic errors \citep{eckart17} are adopted.

This paper is organized as follows. Section~\ref{sec:data} describes observations and data reduction. Section~\ref{sec:analysis} presents details of analysis and results. Implications of our results on the physics of galaxy clusters are discussed in section~\ref{sec:discussion}. Section~\ref{sec:conclusions} summarizes our conclusions. A new redshift measurement of the central galaxy NGC~1275 is presented in appendix~\ref{sec:redshift} and various systematic uncertainties of our results are discussed in appendix~\ref{sec:systematic}. The details of the velocity mapping are shown in appendix~\ref{sec:details}. Throughout the paper, we adopt standard values of cosmological density parameters, $\Omega_{\rm M}=0.3$ and $\Omega_{\Lambda}=0.7$, and the Hubble constant $H_0 = 70$~km~s$^{-1}$~Mpc$^{-1}$. In this cosmology, the angular size of 1~arcmin corresponds to the physical scale of 21~kpc at the updated redshift of NGC~1275, $z=0.017284$. Unless stated otherwise, errors are given at 68\% confidence levels.

\section{Observations and data reduction}
\label{sec:data}

\begin{figure}
 \begin{center}
  \includegraphics[width=8cm]{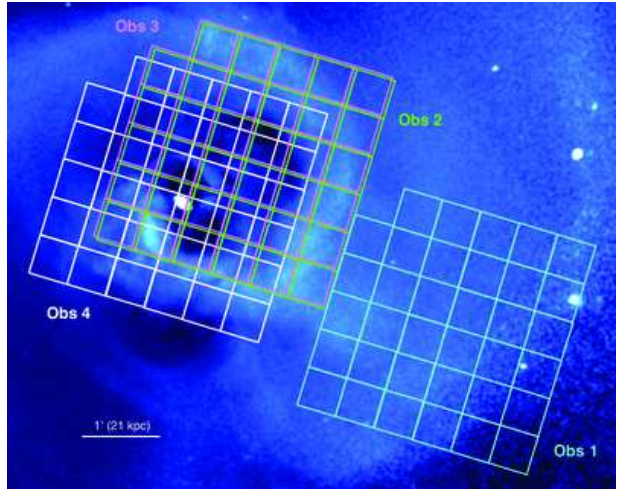}
 \end{center}
\caption{Hitomi SXS pointings of the Perseus cluster performed during the commissioning phase, overlaid on the Chandra 0.5--3.5~keV band relative deviation image \citep[reproduced from][]{Zhuravleva14}. The grids correspond to the 6$\times$6 array of the SXS with a lacking corner for the calibration pixel.}\label{fig:fov}
\end{figure}

\begin{table*}
 \tbl{Summary of the Perseus observations}{%
 \begin{tabular}{llllll}
  \hline
  & ObsID & Observation date & Exposure time (ks) & Pointing direction (RA, Dec) (J2000)\\
  \hline
  Obs~1 & 10040010                     & 2016 February 24 & 48.7  & $(\timeform{3h19m29s.8},  \timeform{+41D29'1''.9})$\\
  Obs~2 & 10040020                     & 2016 February 25 & 97.4  & $(\timeform{3h19m43s.6},  \timeform{+41D31'9''.8})$\\
  Obs~3 & 10040030, 10040040, 10040050 & 2016 March 4     & 146.1 & $(\timeform{3h19m43s.8},  \timeform{+41D31'12''.5})$\\
  Obs~4 & 10040060                     & 2016 March 6     & 45.8  & $(\timeform{3h19m48s.2},  \timeform{+41D30'44''.1})$\\
  \hline
 \end{tabular}}\label{tab:obs}
\end{table*}

The Perseus cluster was observed four times with the SXS during Hitomi's commissioning phase (Obs~1, 2, 3 and 4). A protective gate valve, composed of a $\sim$260~$\mu$m thick beryllium layer, absorbed most X-rays below 2~keV and roughly halved the transmission of X-rays above 2~keV \citep{eckart16}. 
Figure~\ref{fig:fov} shows the footprint of the four pointings superposed on the Chandra 0.5--3.5~keV band relative deviation image \citep[reproduced from][]{Zhuravleva14}. The observations are summarized in table~\ref{tab:obs}. Obs~1 was pointed $\sim$3~arcmin east of the cluster core. Obs~2 and Obs~3, covering the cluster core and centered on NGC~1275, are the only observations analyzed in H16. Obs~4 was pointed $\sim$0.5~arcmin to south-west of the pointing of Obs~2 and Obs~3.

In order to avoid introducing additional systematic uncertainties into our analysis, we have not applied any additional gain correction adopted in other Hitomi Perseus papers \citep[see e.g.][hereafter Atomic~paper]{atomicpaper} unless otherwise quoted. We started the data reduction from the cleaned event list provided by the pipeline processing version 03.01.005.005 \citep{angelini16} with \verb+HEASOFT+ version 6.21. Detailed description of data screening and additional processing steps are described in \citet[][hereafter T~paper]{tpaper} and elsewhere\footnote{ ``The HITOMI Step-By-Step Analysis Guide version 5; https://heasarc.gsfc.nasa.gov/docs/hitomi/analysis/}.

\section{Analysis and results}
\label{sec:analysis}

In this section, we present the analysis and the results subject-by-subject. Several setups are commonly adopted in most of the analyses unless otherwise stated. The atmospheric X-ray emission was modeled as the emission from a single-temperature, thermal plasma in collisional ionization equilibrium attenuated by the Galactic absorption (\verb+TBabs*bapec+). The absorbing hydrogen column density was fixed to the value obtained from Leiden/Argentine/Bonn (LAB) survey \citep[$N_{\rm H}=0.138\times10^{22}~\mathrm{cm}^{-2}$;][]{Kalberla05}. \citet{willingale13} pointed out the effect of the molecular hydrogen column density on the total X-ray absorption, and the effect increases the hydrogen column density by $\sim$50\% in the case of Perseus cluster. We however ignored the correction because (i) we do not use the energy below 1.8~keV, where the effect becomes significant, and (ii) the effect is almost only on the continuum parameters, whose effects are second-order and thus negligible in determining the velocity parameters. We ignored the spectral contributions of the cosmic X-ray background (CXB) as they are negligible compared to the emission of the Perseus cluster \citep{kilbourne16b}. We also ignored the contributions from the non-X-ray background because Hitomi SXS has a significant effective area at high energies \citep{okajima17}, which makes them negligible compared to the X-ray emission components.

We adopted the abundance table of proto-solar metal of \citet{Lodders09} in this paper. Unless otherwise stated, the fitting was performed using {\small XSPEC} v12.9.1 \citep{Arnaud96} with AtomDB v3.0.9 \citep{smith01,foster12}.

The spectra were rebinned so that each energy bin contained at least one event. C-statistics were minimized in the spectral analysis. The redistribution matrix files (RMFs) were generated using the \verb+sxsmkrmf+ tool\footnote{https://heasarc.gsfc.nasa.gov/lheasoft/ftools/headas/sxsmkrmf.html} in which we incorporated the electron loss continuum channel into the redistribution \citep[extra-large-size RMF;][]{leutenegger16}\footnote{For the analyses shown in the main text. We instead used large-size RMFs for the analyses presented in appendices for computational efficiency. The changes in the best-fit values due to the RMF difference are typically less than a few \%.}. Point source ARFs (auxiliary response file) were generated in the 1.8--9.0~keV band using the \verb+aharfgen+ tool\footnote{https://heasarc.gsfc.nasa.gov/ftools/caldb/help/aharfgen.html} at source coordinates (RA, Dec)$=$(\timeform{3h19m48s.1}, \timeform{+41D30'42''}) (J2000).

Hereafter in this paper, we distinguish various kinds of line width using the following notations: $\sigma_\mathrm{v+th}$ is the observed line width with only the instrumental broadening subtracted; $\sigma_\mathrm{v}$ is the line width calculated by subtracting both the thermal broadening ($\sigma_\mathrm{th}$) and the instrumental broadening from the observed line width (i.e., LOS velocity dispersion). Unless stated otherwise, $\sigma_\mathrm{th}$ is computed assuming that electrons and ions have the same temperature. The analysis without this assumption is presented in section~\ref{sec:iontemperature}.

\subsection{Profiles of major emission lines}
\label{sec:spectra}

\begin{figure*}
 \begin{center}
  \includegraphics[width=16cm]{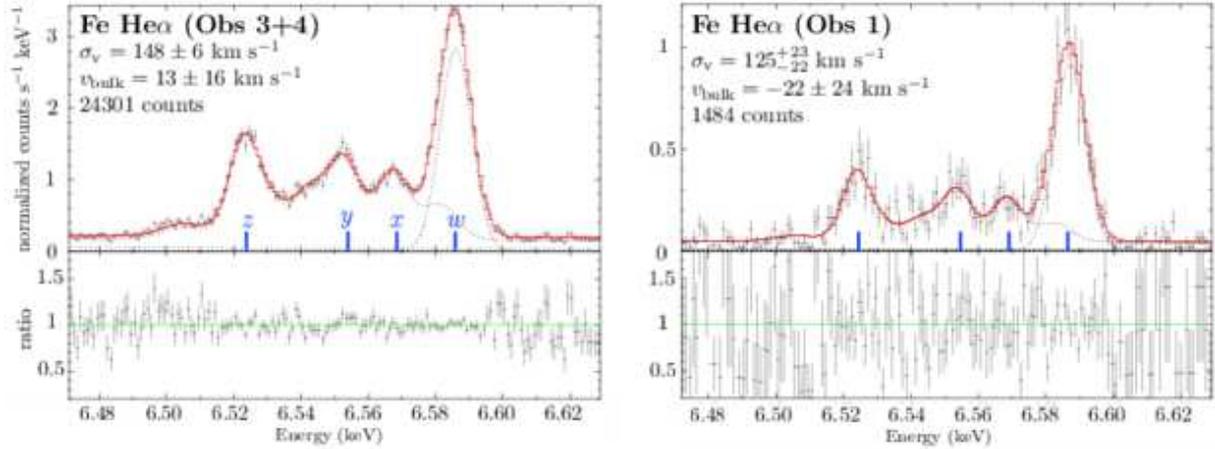}
 \end{center}
 \caption{Fe~He$\alpha$ lines of the full-FOV data of Obs~3+Obs~4 (left) and Obs~1 (right). The LOS velocity dispersion ($\sigma_\mathrm{v}$, {\it w}-line excluded. See also table~\ref{tab:obs34_width}), the bulk velocity calculated with respect to the redshift of NGC~1275 ($v_\mathrm{bulk}$) and the total number of photons in the displayed energy band are shown in each figure. The red curves are the best-fitting models, and the dotted curves are the spectral constituents, i.e., modified APEC or Gaussian. See main text for details. The energy bin size is 1~eV or wider for lower count bins. The resonance line ({\it w}), the intercombination lines ({\it x} and {\it y}), and the forbidden line ({\it z}) are denoted. The letters are as given in \citet{gabriel72}.}
 \label{fig:FeHea}
\end{figure*}

\begin{figure*}
 \begin{center}
  \includegraphics[width=16cm]{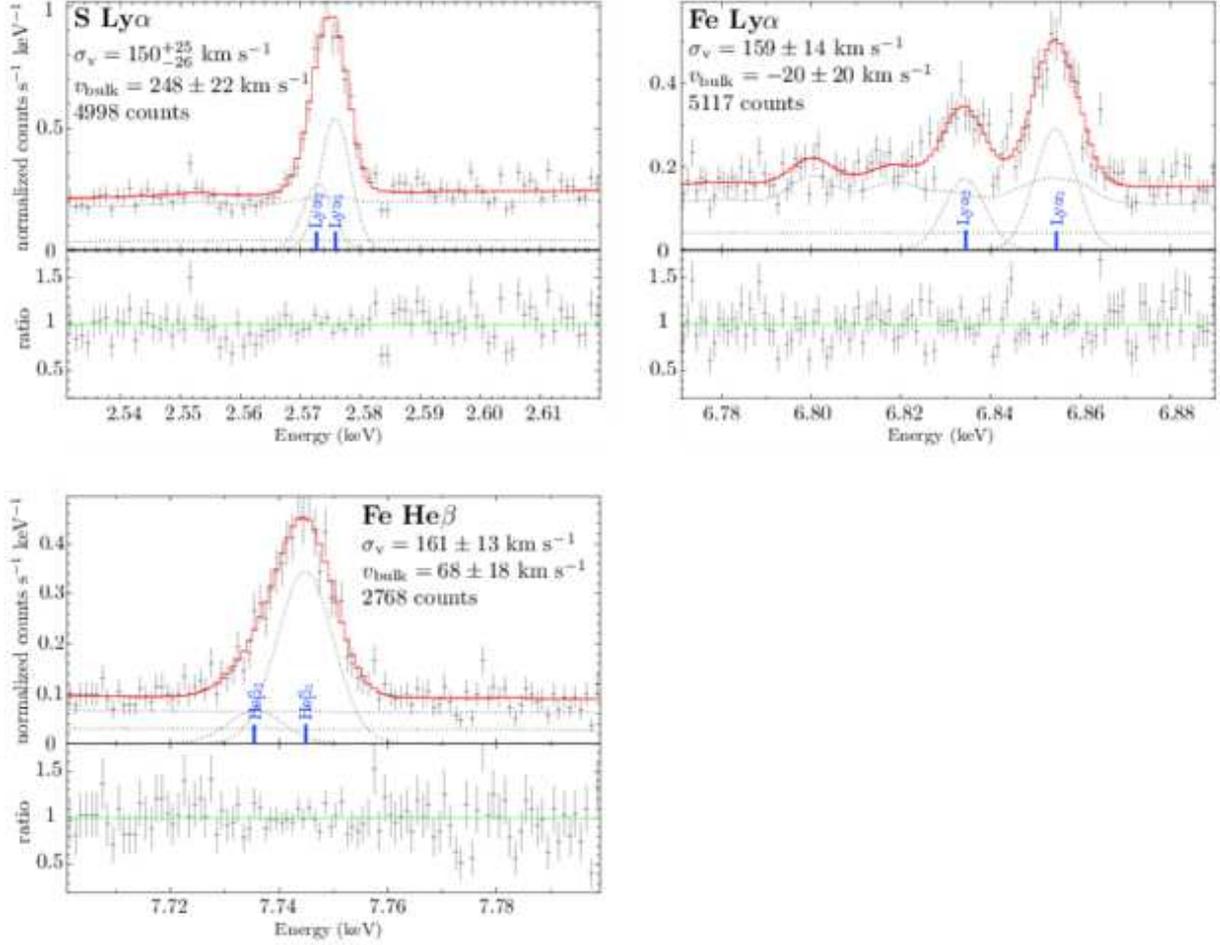} 
 \end{center}
 \caption{Same as figure~\ref{fig:FeHea}, but for S~Ly$\alpha$ (upper left), Fe~Ly$\alpha$ (upper right) and Fe~He$\beta$ (lower left) of Obs~3+4. Representative line are denoted in the figures.}
 \label{fig:otherlines}
\end{figure*}

In this section, we show observed line profiles of bright transitions
and demonstrate qualities of these measurements. The data of Obs~2 were not used
in this section and in section~\ref{sec:velocity}, since Obs~2 (and Obs~1)
contains a previously known systematic uncertainty in the energy
scale, and the almost identical pointing direction to that of Obs~2's is covered by Obs~3.
In figure~\ref{fig:FeHea} we show the Fe~He$\alpha$ emission line
complex from Obs~3+Obs~4, and Obs~1.  The panels in
figure~\ref{fig:otherlines} show S~Ly$\alpha$, Fe~Ly$\alpha$ and
Fe~He$\beta$ lines of Obs~3+Obs~4. The figures indicate the best-fitting
LOS velocity dispersions ($\sigma_\mathrm{v}$) and bulk velocities
calculated with respect to the new stellar absorption line redshift
measurement of NGC~1275 ($v_{\rm bulk}\equiv(z-0.017284)c_0-26.4~{\rm
km~s}^{-1}$, where $c_0$ is the speed of light, $z=0.017284$ is the
redshift of NGC~1275, and $-26.4$~km~s$^{-1}$ is the heliocentric
correction. See also appendix~\ref{sec:redshift} for the redshift
measurement). The net photon count is also indicated.

The best-fitting parameters were obtained as follows: We extracted
spectra from the event file (no additional gain correction applied) for
the entire FOVs of Obs~3, Obs~4, and Obs~1, and we combined the spectra
of Obs~3 and Obs~4. The spectral continua were modeled using a wider
energy band of 1.8--9.0~keV using
\verb+bapec+, and the obtained continuum parameters were used in the subsequent fitting for extracting the parameter values associated with spectral lines performed in narrower energy bands displayed in figures~\ref{fig:FeHea} and \ref{fig:otherlines}. In the \verb+bapec+
modelling, Fe He$\alpha$ {\it w} was manually excluded from the
atomic database and substituted by an external Gaussian, to
minimize the effect of resonance scattering \cite[most pronounced for
Fe He$\alpha$ {\it w}, see][hereafter RS~paper]{rspaper}. In the
spectral line modelling, Fe He$\alpha$ {\it w}, Ly$\alpha_1$ and
Ly$\alpha_2$, He$\beta_1$ and He$\beta_2$, and S Ly$\alpha_1$ and
Ly$\alpha_2$ were manually excluded from the atomic database and
substituted by external Gaussians. For an Fe Ly$\alpha$ feature, the
widths of the two Gaussians were linked to each other, while for Fe
He$\beta$ and S Ly$\alpha$ features, the relative centroid energies and
the relative normalizations of each of the two Gaussians were also fixed
to the database values.

\begin{table*}
 \tbl{LOS velocity dispersions of gas motions, obtained from the Fe~He$\alpha$ line of Obs~3+Obs~4 data.}{%
 \begin{tabular}{lllll}
  \hline
 & Unit & Without {\it $z$-correction} & With {\it $z$-correction}$^*$ \\
\hline
$\sigma_{\mathrm v}$ of {\it w} & (km~s$^{-1}$) & $171^{+4}_{-3}$ & $161\pm3$\\
$\sigma_{\mathrm v}$ excluding {\it w} & (km~s$^{-1}$) & $148\pm6$ & $144\pm6$\\
  \hline
 \end{tabular}}\label{tab:obs34_width}
 \begin{tabnote}
 $^*${\it $z$-correction} is an additional gain alignment among the detector pixels. See also the text.
 \end{tabnote}
\end{table*}

We investigated the effects of the Fe~He$\alpha$ resonance line ({\it w} line) and the energy scale correction on the measured $\sigma_\mathrm{v}$. Table~\ref{tab:obs34_width} shows the LOS velocity dispersion ($\sigma_\mathrm{v}$) measured with or without {\it $z$-correction} -- a rescaling of photon energies for individual SXS pixels in order to force the Fe He-alpha lines align, which has been employed in H16 and \citet{hitomi17} to cancel out most pixel-to-pixel calibration uncertainties, but which also removes any true LOS velocity gradients. The value of $\sigma_\mathrm{v}$ obtained with the {\it w} line is higher than that without the {\it w} line, which provides a hint of resonance scattering (see RS~paper for details).

\subsection{Velocity maps}
\label{sec:velocity}

\begin{figure*}
 \begin{minipage}{0.495\hsize}
  \centering
  \includegraphics[width=8cm]{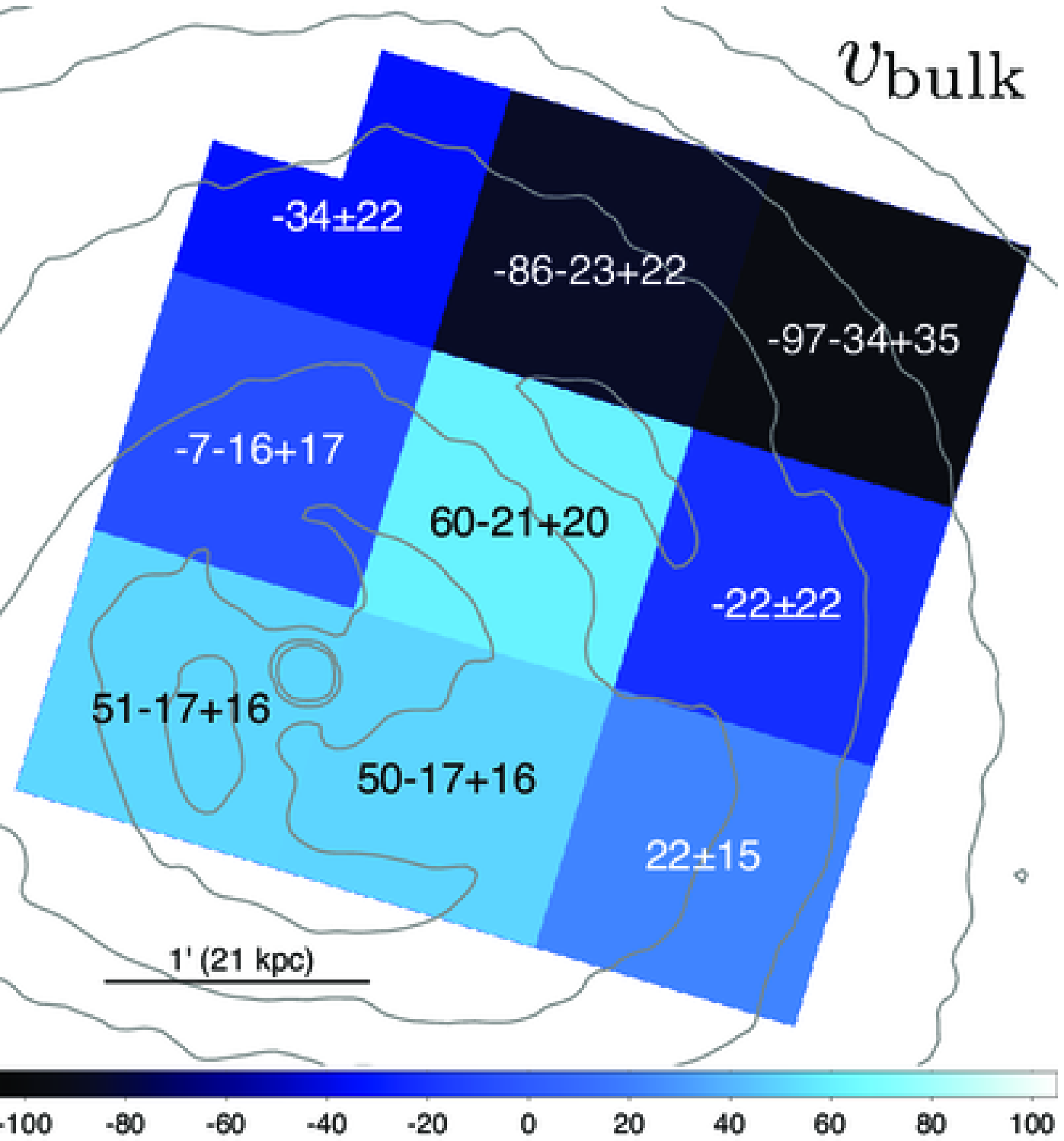}
 \end{minipage}
 \begin{minipage}{0.495\hsize}
  \centering
  \includegraphics[width=8cm]{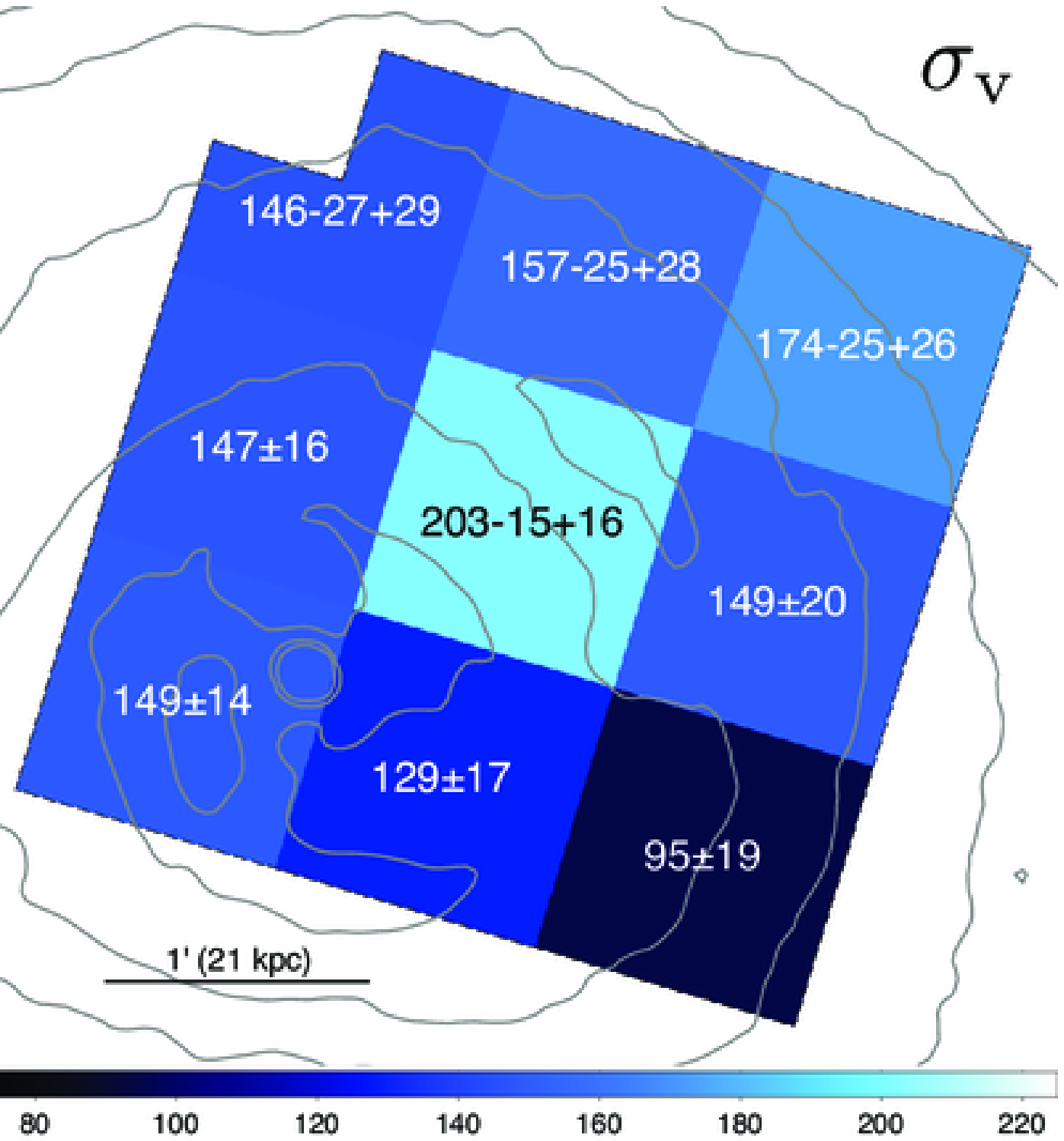}
 \end{minipage}
 \caption{Benchmark velocity maps. {\it Left:} bulk velocity ($v_\mathrm{bulk}$) map with respect to $z=0.017284$ (heliocentric correction of $-26.4~\mathrm{km~s^{-1}}$ applied). {\it Right:} LOS velocity dispersion ($\sigma_\mathrm{v}$) map. The unit of the values is km~s$^{-1}$. Chandra X-ray contours are overlaid. The best-fitting value is overlaid on each region. Only Obs~3 is used and PSF correction is not applied.}\label{fig:comparison}
\end{figure*}

Firstly, we extracted the benchmark velocity maps by objectively dividing the 6~pixel$\times$6~pixel array into 9 subarrays of 2$\times$2~pixels and fitted the spectrum of each region independently, in order to compare the effects of the difference in software and data pipeline versions between H16 and this paper. All model parameters apart from the hydrogen column density were allowed to vary. Only Obs~3 was used for the benchmark maps and the fitting was done using a narrow energy range of 6.4--6.7~keV, excluding the energy band corresponding to the resonance line of Fe He$\alpha$ in the observer-frame (6.575--6.6~keV) to avoid the systematics originating from the possible line broadening due to the resonant scattering effect. Figure~\ref{fig:comparison} left shows the bulk velocity ($v_\mathrm{bulk}$) map with respect to $z=0.017284$ (heliocentric correction of $-26.4~\mathrm{km~s^{-1}}$ applied) and figure~\ref{fig:comparison} right shows the LOS velocity dispersion ($\sigma_\mathrm{v}$) map. We found a similar trend to the H16 results.

\begin{figure*}
 \begin{minipage}{0.495\hsize}
  \centering
  \includegraphics[width=8.0cm]{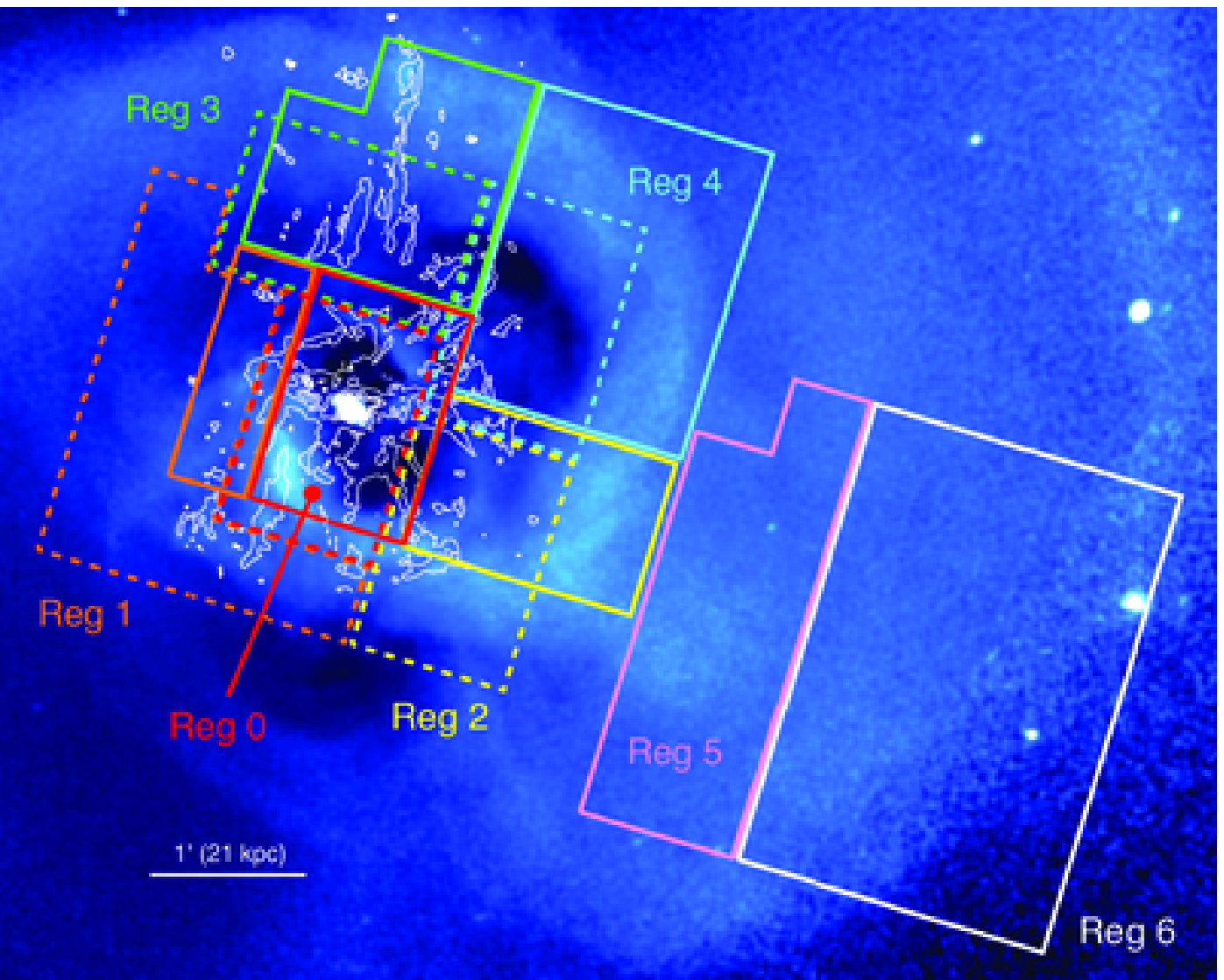}
 \end{minipage}
 \begin{minipage}{0.495\hsize}
  \centering
  \includegraphics[width=8.0cm]{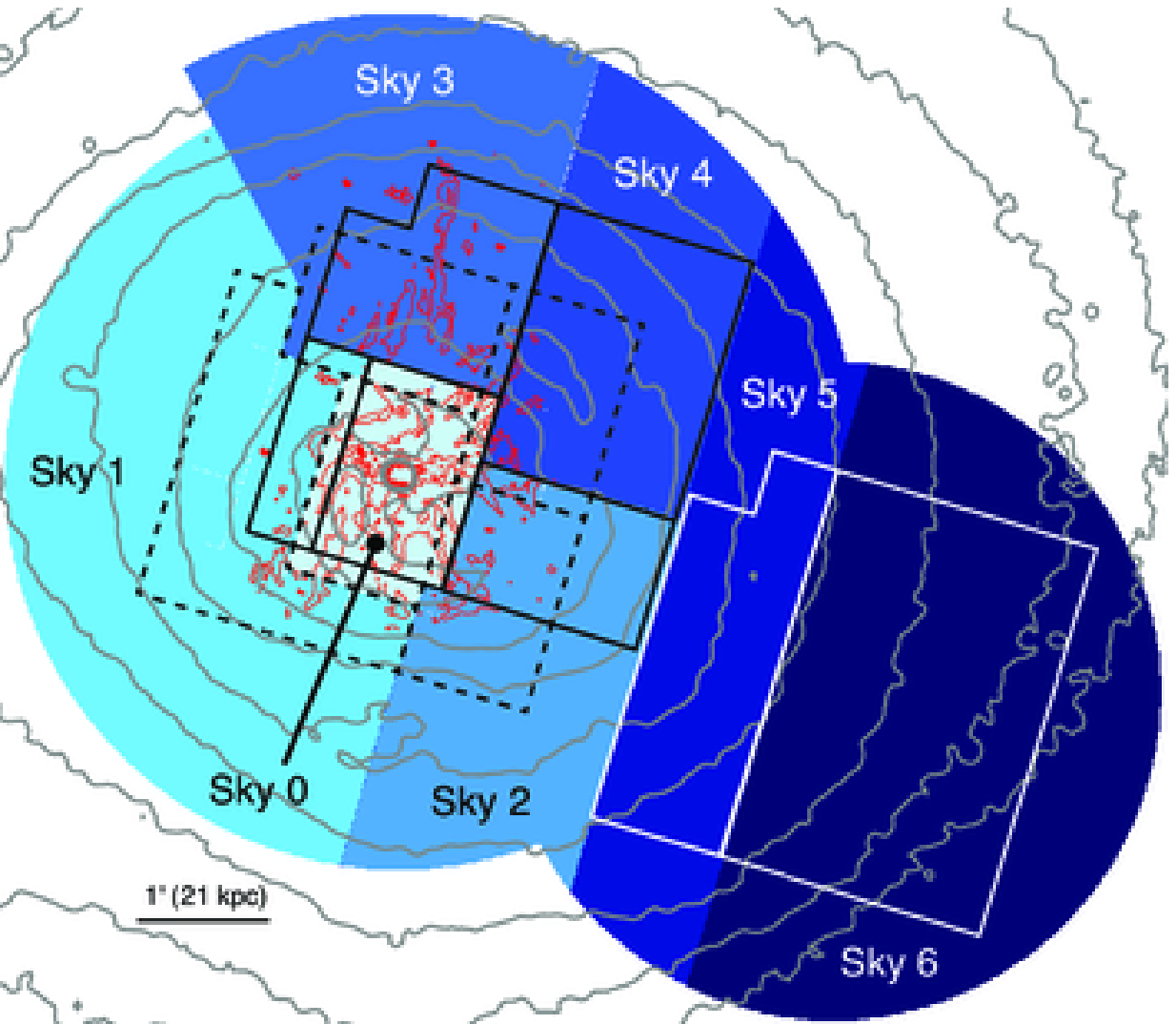}
 \end{minipage}
 \caption{The regions used for the velocity mapping. {\it Left:} distinct regions defined by discrete pixels are identified by color coding and number and overlaid on the Chandra relative deviation image. {\it Right:} the corresponding regions when PSF is taken into account. The Chandra X-ray contours are overlaid. H$\alpha$ contours \citep{conselice01} are also overlaid in white (left) or red (right). The solid-lined polygons are the regions associated with Obs~1 or Obs~3, and the dashed-lined polygons are the regions associated with Obs~4. See also figure~\ref{fig:fov}.}\label{fig:region}
\end{figure*}

Secondly, we extracted the velocity maps from the regions associated with physically interesting phenomena. Figure~\ref{fig:region} shows the regions used for the velocity mapping. Most of the regions correspond to a specific feature pointed out in the literature \citep[e.g.][]{Churazov00,Fabian06,salome2011}: Reg~0 represents the central AGN and the cluster core; Reg~3 covers the northern filaments; and Reg~4 surrounds the northwestern ghost bubble. We excluded Obs~2 in our velocity mapping to avoid potential systematic uncertainties (see appendix~\ref{sec:gain} for details).

The PSF of the telescope (1.2~arcmin HPD) is rather broad, and thus X-ray photons are scattered out of the FOV and into adjacent regions. Also conversely, photons from outside the detector array's footprint are scattered into the array.

In order to account for the scattering from outside the detector array's footprint, we extended the sky areas for Reg~1 and Reg~2 to a radius of $r=3$~arcmin from the central AGN. We extended Reg~3, 4, and 5 to a radius of 3.5~arcmin from the central AGN. Reg~5 and 6 were likewise extended to a radius of 2.5~arcmin from the center of the FOV of Obs~1. Reg~2 included a part of the region of the $r < 2.5$~arcmin circle and Reg~5 also included a part of the region of the $r < 3.5$~arcmin circle. Sky regions are shown in the right panel of figure~\ref{fig:region}. As the level of PSF blending from outside these regions was found to be less than 1~\%, we ignored them. We assumed a uniform plasma properties within each sky region.

\begin{table*}
\tbl{Ratio of PSF blending effect on each integration region in the 6.4--6.7 keV band in units of percent.}{%
\begin{tabular}{ccccccccc}
\hline
& & \multicolumn{7}{c}{Sky region}  \\
& & Sky~0 & Sky~1 & Sky~2 & Sky~3 & Sky~4 & Sky~5 & Sky~6 \\
\hline
\multirow{12}*{\rotatebox[origin=c]{90}{Integration region}}
& Reg~0 Obs~3 & 62.3 & 10.1 & 13.8 & 7.4  & 6.1  & 0.4  & 0.1 \\ 
& Reg~0 Obs~4 & 64.2 & 16.6 & 10.2 & 5.4  & 3.2  & 0.3  & 0.1 \\
& Reg~1 Obs~3 & 43.9 & 43.3 & 3.0  & 8.3  & 1.2  & 0.2  & 0.1 \\
& Reg~1 Obs~4 & 22.1 & 67.2 & 4.3  & 5.5  & 0.7  & 0.2  & 0.1 \\
& Reg~2 Obs~3 & 10.2 & 2.8  & 65.5 & 1.5  & 12.0 & 7.6  & 0.5 \\
& Reg~2 Obs~4 & 17.8 & 6.5  & 66.5 & 1.5  & 5.7  & 1.9  & 0.2 \\
& Reg~3 Obs~3 & 12.7 & 6.8  & 2.5  & 63.6 & 13.9 & 0.5  & 0.1 \\
& Reg~3 Obs~4 & 22.7 & 15.7 & 2.9  & 51.3 & 7.0  & 0.3  & 0.1 \\
& Reg~4 Obs~3 & 8.2  & 1.8  & 12.6 & 8.5  & 61.5 & 6.8  & 0.5 \\
& Reg~4 Obs~4 & 17.5 & 2.4  & 16.4 & 12.6 & 48.9 & 2.0  & 0.2 \\
& Reg~5 Obs~1 & 1.3  & 0.9  & 17.5 & 0.4  & 4.0  & 60.8 & 15.0 \\
& Reg~6 Obs~1 & 0.8  & 0.8  & 4.4  & 0.4  & 1.6  & 16.0 & 75.9 \\
 \hline
 \end{tabular}}\label{tab:psf}
\begin{tabnote}
Sky regions correspond to the regions shown in the right panel of figure~\ref{fig:region} and integration regions are associated with the regions indicated in the left panel of figure~\ref{fig:region}. The fractions of photons coming from each sky region to one integration region appear in the same row. The level of PSF blending from outside these regions was found to be less than 1~\% and not listed in the table. For example, Reg~1 Obs~3 is strongly affected by scattered photons from Sky~0, and the contamination from Sky~0 to Reg~5 or Reg~6 is almost zero.
\end{tabnote}
\end{table*}

In order to model all the spectra simultaneously, we estimated the relative flux contributions from all the sky regions (figure~\ref{fig:region} right) to every single integration region (figure~\ref{fig:region} left). We measured the quantity of PSF scattering from inside or outside the corresponding sky using \verb+aharfgen+. For the input, we used the deep Chandra image in the broad band of 1.8--9.0~keV and an image in the 6.4--6.7~keV including the line emission only (see appendix~\ref{sec:details}). We show a matrix of its effect in the 6.4--6.7 keV band in table~\ref{tab:psf}. We also checked its effect in the 1.8--9.0 keV band. The trend in the 1.8--9.0 keV band is consistent with that in the 6.4--6.7 keV band.

In order to determine ICM velocities, we fitted spectra from all regions simultaneously, taking scattering into account (see appendix~\ref{sec:xcm} for technical details). We first obtained the PSF-corrected values of the temperature, Fe abundance and normalization of each region. This fitting was done in the energy range of 1.8--9.0~keV, excluding the narrow energy range of 6.4--6.7~keV, and the AGN contribution to the spectra was included using the model shown in \citet[][hereafter AGN~paper]{agnpaper}, after convolution with the point source ARFs. The velocity width and redshift of each plasma model were fixed to 160~km~s$^{-1}$ and 0.017284 respectively. The obtained C-statistic/d.o.f. (degree of freedom) in the continuum fitting is 63146.77/68003. Detailed description of the measurement of the continuum parameters are shown in AGN~paper and T~paper.

After determining the self-consistent parameter set of the continuum as mentioned above, we again fitted all the spectra simultaneously to obtain the parameters associated with spectral lines. This time, the temperatures and normalizations were fixed to the above obtained values, and the Fe abundance, the LOS velocity dispersion and the redshift were allowed to vary. The fitting was done using a narrow energy range of 6.4--6.7~keV, excluding the energy band corresponding to the resonance line in the observer-frame (6.575--6.6~keV). The obtained C-statistic/d.o.f. in the velocity fitting is 2822.38/2896.

\begin{figure*}
 \begin{minipage}{0.495\hsize}
  \centering
  \includegraphics[width=7.5cm]{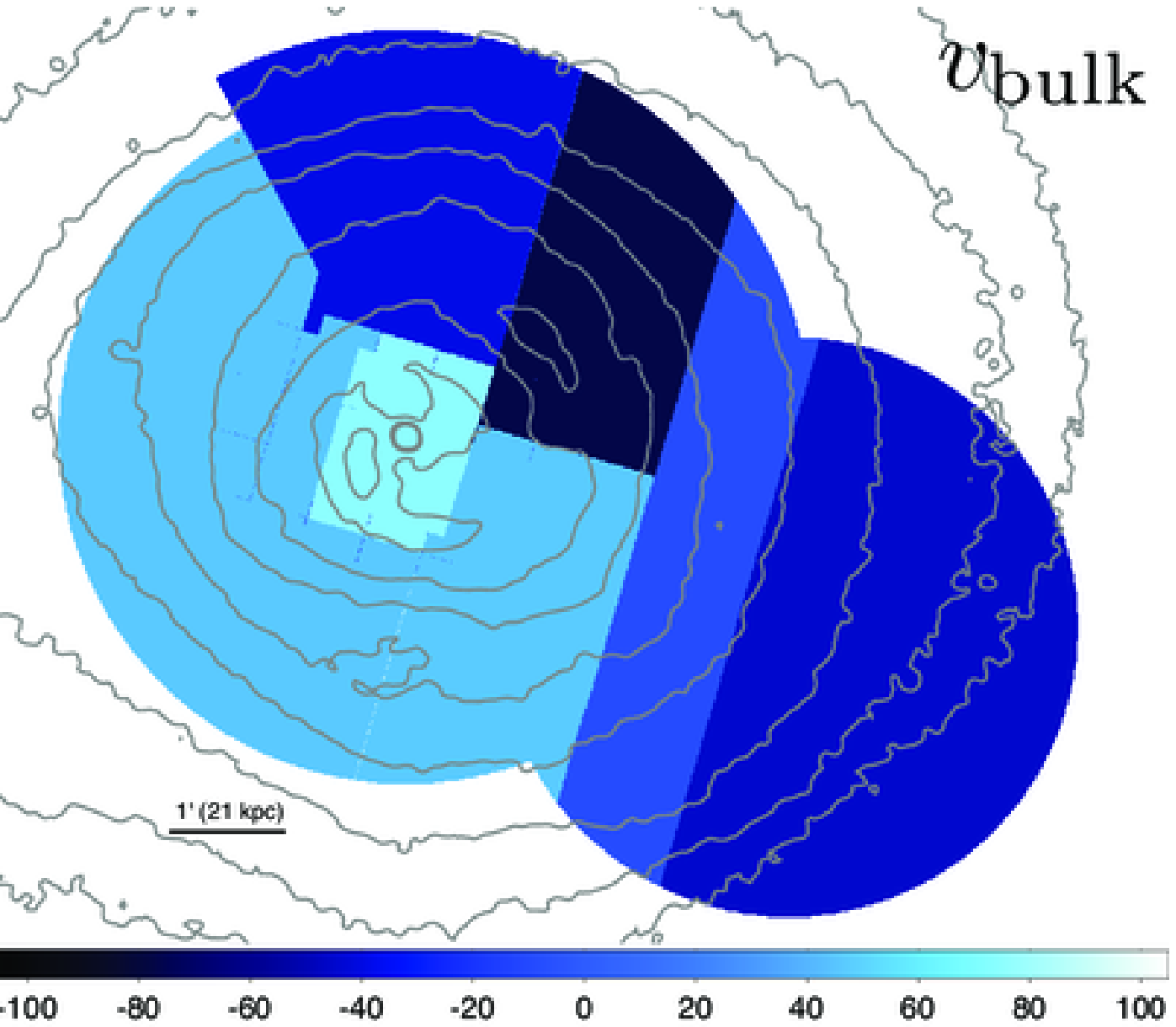}
 \end{minipage}
 \begin{minipage}{0.495\hsize}
  \centering
  \includegraphics[width=7.5cm]{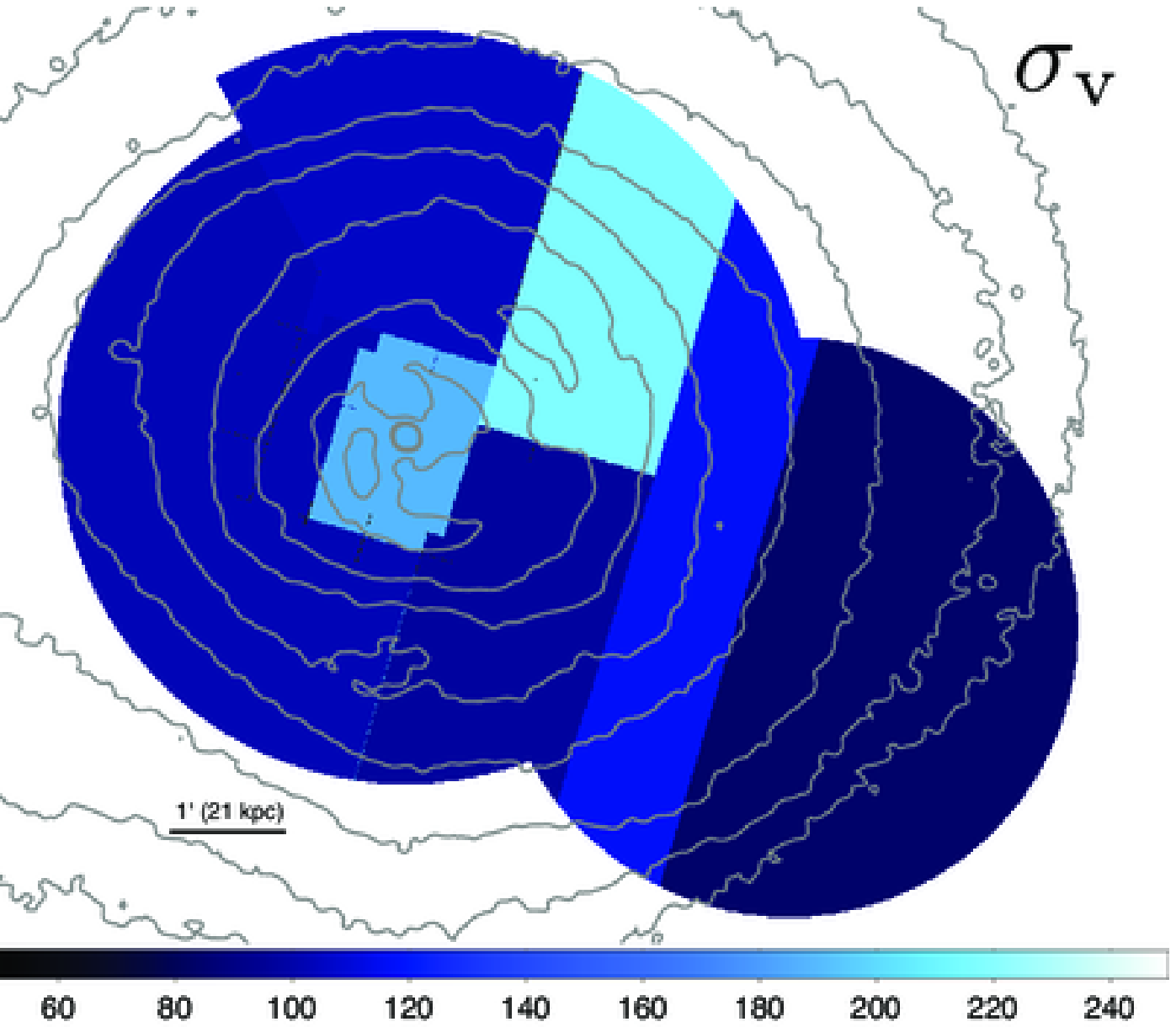}
 \end{minipage}
 \caption{{\it Left:} PSF corrected bulk velocity ($v_\mathrm{bulk}$) map with respect to $z=0.017284$ (heliocentric correction applied). {\it Right:} PSF corrected LOS velocity dispersion ($\sigma_\mathrm{v}$) map. The unit of the values is km~s$^{-1}$. The Chandra X-ray contours are overlaid.}\label{fig:velocity_psfcor}
\end{figure*}

\begin{figure*}
 \begin{minipage}{0.495\hsize}
  \centering
  \includegraphics[width=7.5cm]{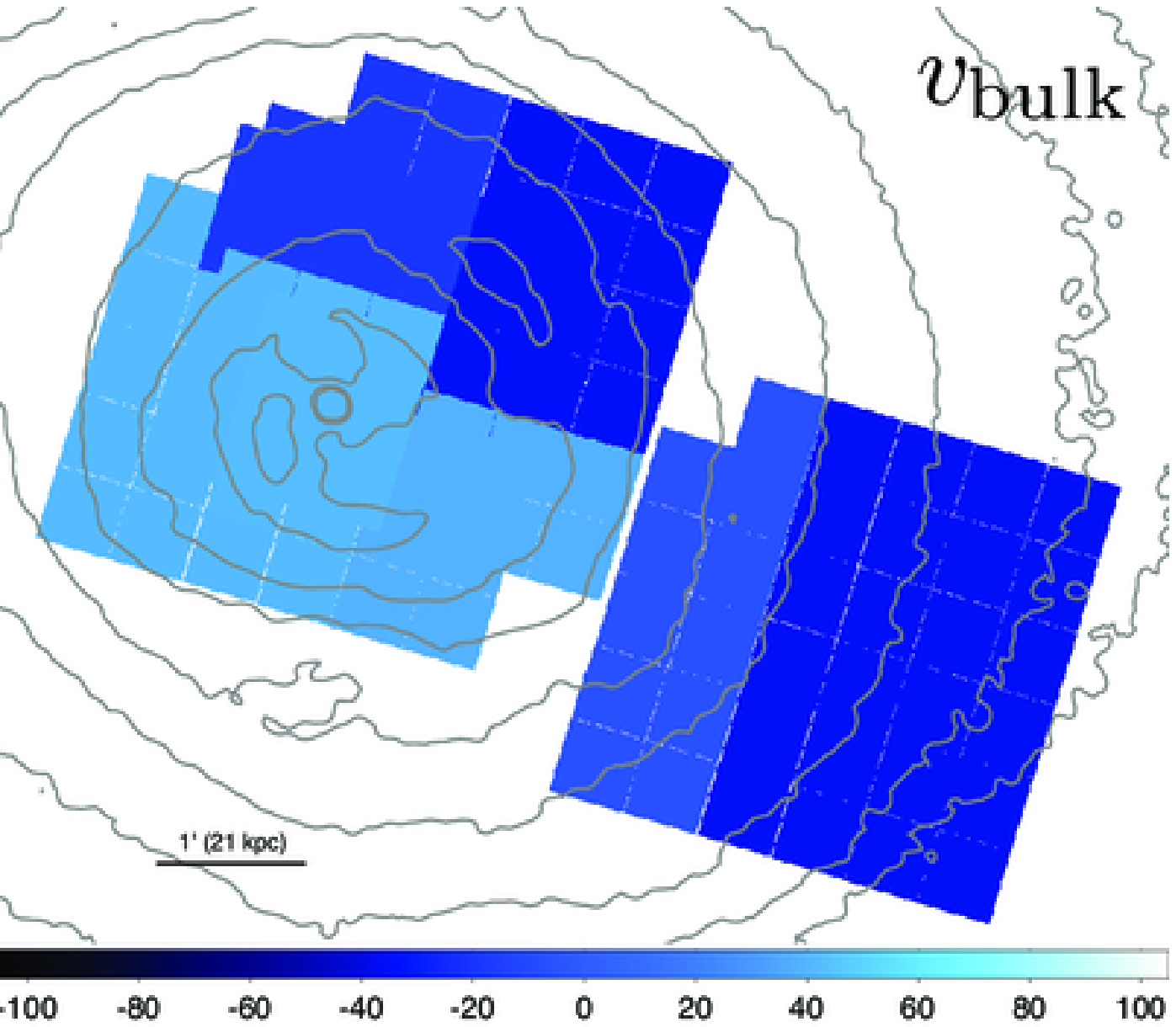}
 \end{minipage}
 \begin{minipage}{0.495\hsize}
  \centering
  \includegraphics[width=7.5cm]{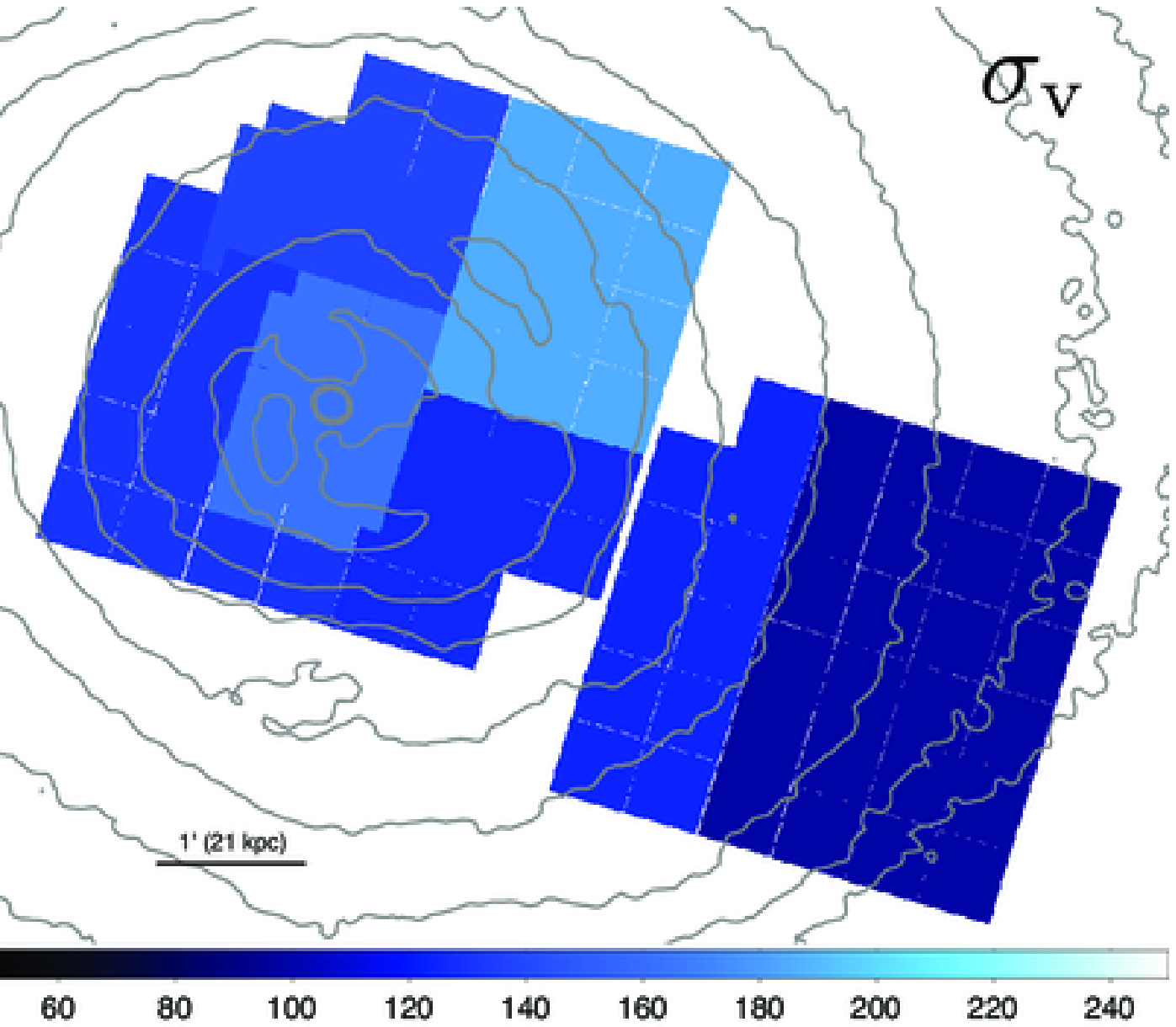}
 \end{minipage}
 \caption{Same as figure~\ref{fig:velocity_psfcor}, but PSF correction is not applied.}\label{fig:velocity_nocor}
\end{figure*}

\begin{table*}
  \tbl{Best-fitting bulk velocity ($v_\mathrm{bulk}$) and LOS velocity dispersion ($\sigma_\mathrm{v}$) values of with and without PSF correction.}{%
  \begin{tabular}{lllll}
   \hline
   & \multicolumn{2}{c}{PSF corrected} & \multicolumn{2}{c}{PSF uncorrected}\\
  Region & $v_\mathrm{bulk}$ (km~s$^{-1}$) & $\sigma_\mathrm{v}$ (km~s$^{-1}$) & $v_\mathrm{bulk}$ (km~s$^{-1}$) & $\sigma_\mathrm{v}$ (km~s$^{-1}$)\\
  \hline
  Reg~0 & $75_{-28}^{+26}$  & $189_{-18}^{+19}$ & $43_{-13}^{+12}$  & $163_{-10}^{+10}$  \\
  Reg~1 & $46_{-19}^{+19}$  & $103_{-20}^{+19}$ & $42_{-12}^{+12}$  & $131_{-11}^{+11}$  \\
  Reg~2 & $47_{-14}^{+14}$  & $98_{-17}^{+17}$  & $39_{-11}^{+11}$  & $126_{-12}^{+12}$  \\
  Reg~3 & $-39_{-16}^{+15}$ & $106_{-20}^{+20}$ & $-19_{-11}^{+11}$ & $138_{-12}^{+12}$  \\
  Reg~4 & $-77_{-28}^{+29}$ & $218_{-21}^{+21}$ & $-35_{-14}^{+15}$ & $186_{-12}^{+12}$  \\
  Reg~5 & $-9_{-56}^{+55}$  & $117_{-73}^{+62}$ & $-6_{-26}^{+25}$  & $125_{-28}^{+28}$  \\
  Reg~6 & $-45_{-29}^{+29}$ & $84_{-54}^{+44}$  & $-35_{-22}^{+22}$ & $99_{-32}^{+31}$  \\
     \hline
  \end{tabular}}\label{tab:velocity}
\end{table*}

Figure~\ref{fig:velocity_psfcor} shows the obtained velocity maps with PSF correction. The corresponding velocity maps without PSF correction are shown in figure~\ref{fig:velocity_nocor} for comparison. The best-fitting values are listed in table~\ref{tab:velocity}. The heliocentric correction of $-26.4~\mathrm{km~s^{-1}}$ is applied in the bulk velocity maps.

When producing the PSF-corrected maps, the twelve spectra (Obs~3 and Obs~4 for Reg~0 to Reg~4 and Obs~1 for Reg~5 and Reg~6) were fitted simultaneously with all the cross-terms being incorporated through the matrix shown in table~\ref{tab:psf}. The fitting procedure is complex and deconvolution is often unstable. We thus carefully examined the robustness of the results. These included the check of two parameter confidence surfaces based on C-statistics, i.e., redshift vs LOS velocity dispersion, Fe abundance vs redshift, and Fe abundance vs LOS velocity dispersion for each region, and LOS velocity dispersion vs LOS velocity dispersion and redshift vs redshift for each combination of regions. The redshift, LOS velocity dispersion, and Fe abundance are within 0.0165--0.0180, 0.0--250~km~s$^{-1}$, and 0.35--0.85~solar, respectively. We found no strong correlations among parameters and also confirmed that the true minimum was found in the fitting.

In appendix~\ref{sec:details}, we also desrcibe a different method of deriving the velocties that uses only the {\it w} line (which has been excluded in the fit above). It gives qualitatively similar results with the expected higher values of velocity dispersion. Further detailed investigations of the systematic uncertainties and various checks of the results are presented in appendices~\ref{sec:systematic} and \ref{sec:details}.

\subsection{Limits on non-Gaussianity of line shapes}
\label{sec:nongaussianity}

As shown in section~\ref{sec:spectra}, the observed widths of the Fe lines ($\sigma\sim 4$~eV) are much broader than those expected by the convolution of the line spread function of the SXS (FWHM $\sim5$~eV or $\sigma\sim2$~eV) with the thermal width ($\sigma_{\rm th}\sim2$~eV for Fe at $kT\sim4$~keV). Note also that uncertainties of instrumental energy scale and the line spread function at around 6~keV are smaller than the observed widths, as shown in appendix~\ref{sec:systematic}. They are instead governed by hydrodynamic motion of the gas. We thus aim to obtain further information on the gas velocity distribution by examining the line shapes in detail. In figures~\ref{fig:FeHea} and \ref{fig:otherlines}, fitting results of S Ly$\alpha$, Fe He$\alpha$, Ly$\alpha$, and He$\beta$ lines from Obs~3 and 4 are shown with residuals (ratios of the data to the best-fit model). In what follows, we make use of Obs~2 to improve the statistics and further investigate the line shapes.

\begin{table*}
\tbl{Centroid energy in the observer frame, width, significance, and goodness-of-fit of lines detected at $> 5\sigma$.}{
\begin{tabular}{@{\extracolsep{3pt}}cccccccc@{}}
\hline
     & \multicolumn{3}{c}{Line information} & \multicolumn{3}{c}{Fitting information$^*$} & \\\cline{2-4} \cline{5-7}
Line & Centroid energy$^\dagger$ & $\sigma_{\rm v+th}$ & Significance$^\ddagger$ & Energy band & C-statistic & d.o.f. & Note$^{**}$\\
     & (eV)   & (km\,s$^{-1}$)  &                  & (keV)       &             &        &\\
\hline
Si Ly$\alpha$ & $1969.32\pm 0.21$ & $224^{+49}_{-54}$ & 12.9 & 1.945--1.995 & 40.27 & 45&(1)\\
Si Ly$\beta$ & $2333.73\pm 0.49$ & $327^{+71}_{-68}$ & 7.4 & 2.28--2.38 & 71.66 & 94&\\
S He$\alpha$ & $2417.05\pm 0.38$ & $256^{+59}_{-57}$ & 8.1 & 2.355--2.45 & 73.47 & 90&\\
S Ly$\alpha$ & $2575.83\pm 0.11$ & $192^{+21}_{-22}$ & 27.2 & 2.53--2.62 & 117.17 & 85&\\
S Ly$\beta$ & $3052.33\pm 0.26$ & $198^{+39}_{-38}$ & 10.9 & 3.00--3.14 & 116.85 & 132&\\
Ar He$\alpha$ & $3084.46\pm 0.34$ & $150^{+47}_{-50}$ & 8.5 & 3.00--3.14 & 99.95 & 132&\\
Ar Ly$\alpha$ & $3265.12\pm 0.27$ & $260^{+38}_{-37}$ & 14.3 & 3.235--3.29 & 38.94 & 50&(2)\\
Ca He$\alpha$ & $3835.26\pm 0.19$ & $186^{+21}_{-20}$ & 15.8 & 3.77--3.855 & 55.85 & 79&\\
Ca Ly$\alpha$ & $4036.97\pm 0.35$ & $202^{+39}_{-33}$ & 13.4 & 3.98--4.10$^\S$ & 94.01 & 95&\\
Fe He$\alpha$ $z$ & $6522.97\pm 0.11$ & $166\pm  5$ & 44.3 & 6.47--6.63 $^\|$& 167.79 & 148&\\
Fe He$\alpha$ $w$ & $6586.13^{+0.06}_{-0.07}$ & $195\pm  3$ & 78.8 & 6.47--6.63$^\|$ & 182.37 & 148&(3)\\
Fe Ly$\alpha$ & $6854.49\pm 0.24$ & $183\pm 11$ & 18.1 & 6.77--6.89 & 143.74 & 113&\\
Ni He$\alpha$ & $7671.73^{+0.60}_{-0.61}$ & $224^{+36}_{-33}$ & 8.0 & 7.55--7.71 & 145.35 & 155&(4)\\
Fe He$\beta$ & $7744.83^{+0.22}_{-0.23}$ & $178^{+11}_{-10}$ & 27.5 & 7.70--7.80 & 82.35 & 94&\\
Fe Ly$\beta$ & $8112.19^{+0.84}_{-0.46}$ & $  0^{+75}_{- 0}$ & 5.9 & 8.05--8.22 & 152.86 & 162&(5)\\
Fe He$\gamma$ & $8152.44\pm 0.50$ & $189\pm 20$ & 12.5 & 8.05--8.22 & 146.75 & 162&(6)\\
\hline
\end{tabular}}\label{tab:width_obs234}
\begin{tabnote}
$^*$ C-statistic and d.o.f. are those in the specified energy band.\\
$^\dagger$ Energy of the most prominent component, unless specified otherwise.\\
$^\ddagger$ Significance was determined by dividing the normalization by its $1\sigma$ error.\\
$^\S$ Energy range from 4.07~keV to 4.09~keV was ignored, to exclude Ar Ly$\gamma$.\\
$^\|$ Gaussians were used for both $z$ and $w$ lines.\\ 
$^{**}$ (1) Line width changed from $1.85_{-0.42}^{+0.41}$~eV to $1.50_{-0.36}^{+0.33}$~eV, by adding Obs 2 data. The parameters may be unreliable. (2) Line width changed from $2.24_{-0.52}^{+0.51}$~eV to $2.88\pm0.42$~eV, by adding Obs 2 data. The parameters may be unreliable. (3) This line is likely to be optically thick and affected by resonance scattering. (4) This energy range is contaminated by Fe satellite lines, and the parameters may be unreliable. (5) This energy range is contaminated by various satellite lines. In addition, the line width changed from $9.0_{-2.6}^{+2.8}$~eV to $0.0_{-0.0}^{+2.1}$~eV by adding Obs 2 data. The parameters may be unreliable. (6) This energy range is contaminated by various satellite lines. The parameters might be affected by them.
\end{tabnote}
\end{table*}


The observed centroid energy of the Fe He$\alpha$ resonance line
of Obs~2 is about 1.8~eV lower than that of Obs~3, and its width
($\sigma$) is about 0.36~eV broader, despite their similar pointing
directions. Obs~2 (and Obs~1) occurred while the SXS dewar was still
coming into thermal equilibrium after launch \citep{fujimoto16}, and
these discrepancies come from the limitations of the method used to
correct the drifting energy scale. The energy scale of the Obs~2 data
was corrected as follows, to align their line centers. First, the
centroid energy of each line of Obs~2 and 3 was determined by fitting
the data separately. Then the energy (PI column) of each photon in the
event file of Obs~2 was recalculated by multiplying a factor $E_{\rm
Obs~3}/E_{\rm Obs~2}$, where $E_{\rm Obs~2}$ and $E_{\rm Obs~3}$ are the
best-fit line center energies of Obs~2 and Obs~3, respectively. The
event files of Obs~2, 3, and 4 were then merged and spectral files were
generated.  Note that the correction factor was determined for each line
and hence, a spectral file was generated for each line separately. Note
also that no additional gain alignment among the detector pixels was
applied. The spectra were fitted in the same manner as described in
section~\ref{sec:spectra}. Note that, for Fe He$\alpha$, the resonance
({\it w}) line and the forbidden ({\it z}) line were manually excluded
from the atomic database and substituded by external Gaussians, to
determine the parameters of these lines. The fitting results are shown
in table~\ref{tab:width_obs234}.

\begin{figure*}
\begin{center}
\includegraphics[width=16.5cm]{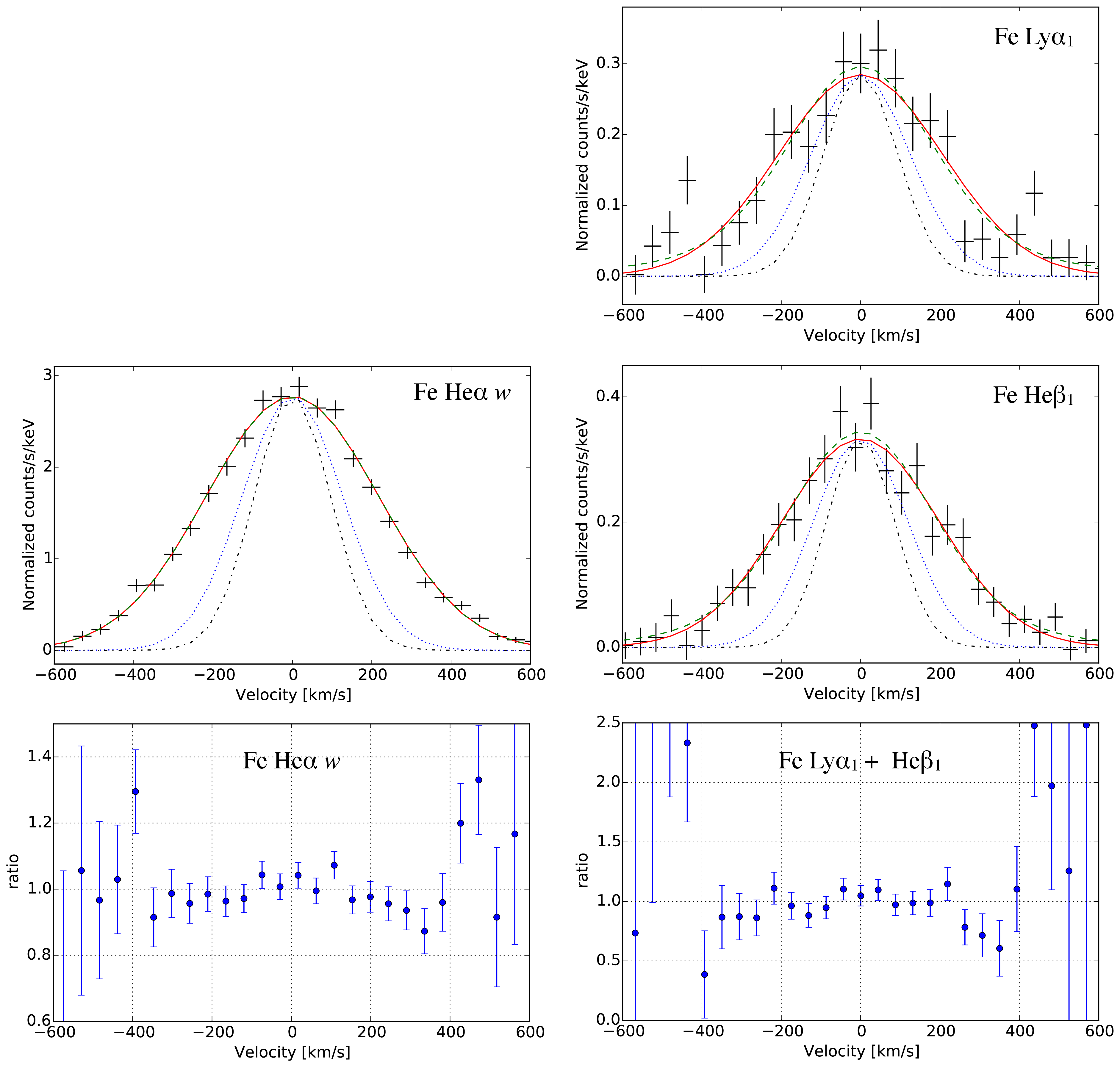}
\end{center}
 \caption{(Upper panels) Data and best-fit models of Fe He$\alpha$ {\it w}, Ly$\alpha_1$, and He$\beta_1$. The continuum model and the components other than the main line were subtracted. Solid (red) and dashed (green) lines represent the best-fit Gaussian and Voigtian profiles, respectively. Instrumental broadening with and without thermal broadening are indicated with dotted (blue) and dashed-dotted (black) lines. The horizontal axis is the velocity converted from the observed energy, where the line center is set at the origin. The bin size is 1~eV in the energy space, which corresponds to 45.5~km\,s$^{-1}$, 43.7~km\,s$^{-1}$, and 38.7~km\,s$^{-1}$, respectively. (Lower panels) The ratio spectra of the data to the best-fit Gaussian models, (left) for Fe He$\alpha$ {\it w}, and (right) for Fe Ly$\alpha_1$ and He$\beta_1$ co-added. Note that the line spread function is not deconvolved from the data.}
 \label{fig:stacked_ratio_spectra}
\end{figure*}

In this section, we focus on three brightest and less contaminated Fe transitions, He$\alpha$, Ly$\alpha$, and He$\beta$. They are from the single element and have a common thermal broadening. In addition, their energies are close enough that we can assume no significant difference in the detector line spread functions. Any astronomical velocity deviation components can cause common residuals of the line shapes in velocity space. Figure~\ref{fig:stacked_ratio_spectra} shows the spectra of these lines in velocity space, after subtracting the best-fitting continuum model and the components other than the main line (He$\alpha$ {\it w}, Ly$\alpha_1$, and He$\beta_1$), where the line center energies were set at the origin of the velocity. As we are interested in deviations from Gaussianity, ratios of the data to the best-fit Gaussian models were also shown in figure~\ref{fig:stacked_ratio_spectra}. Ratios of Ly$\alpha_1$ and He$\beta_1$ were co-added. Positive (ratio $>1$) features are seen at around $\pm(400$--$500)$~km\,s$^{-1}$, while there is a negative (ratio $<1$) feature at around $+300$~km\,s$^{-1}$. However, they are not as broad as the detector line spread function (FWHM $\sim230$~km\,s$^{-1}$). Therefore, we do not conclude that these are cluster-related velocity structures.

\begin{table*}
 \tbl{Best-fit widths when Voigt functions were used.}{
  \begin{tabular}{cccccc}
\hline
 & Gaussian width ($\sigma$) & Lorentzian width (FWHM) & C-statistic & d.o.f. & Natural width (FWHM)$^*$\\
 & (km\,s$^{-1}$) & (km\,s$^{-1}$) & & & (km\,s$^{-1}$)\\
\hline
Fe He$\alpha$ {\it w} & $194\pm 3$ & $0.10^{+0.09}_{-0.03}$ & 182.04 & 147 & 13.9\\
Fe Ly$\alpha$ & $113^{+14}_{-13}$ &$172^{+29}_{-17}$ & 137.44 & 112 & 8.2\\
Fe He$\beta$ & $137\pm11$ & $114^{+20}_{-19}$ & 80.39 & 93 & 3.0\\
\hline
\end{tabular}}\label{tab:width_obs234_Voigt}
\begin{tabnote}
$^*$ Calculated using the Einstein $A$ coefficient shown in AtomDB.\\
\end{tabnote}
\end{table*}

We also fitted each line in the same manner as described above, but
using Voigt functions\footnote{For the Voigt function fitting, we used
the patched model that is the same code as implemented in XSPEC
12.9.1l. See also
https://heasarc.gsfc.nasa.gov/xanadu/xspec/issues/issues.html.} instead
of Gaussians, for He$\alpha$ {\it w}, Ly$\alpha_1$, Ly$\alpha_2$,
He$\beta_1$, and He$\beta_2$. The best-fitting shapes after subtracting
the continuum and the components other than the main line are shown with
dashed curves in the upper panels of
figure~\ref{fig:stacked_ratio_spectra}, and the best-fit widths are
summarized in table~\ref{tab:width_obs234_Voigt}. The Lorentzian widths
of Ly$\alpha$ and He$\beta$ were much broader than the natural
width. This may be due to large positive deviations at around
$\pm(400$--$500)$~km\,s$^{-1}$. On the other hand, it was smaller
than the natural width for He$\alpha$.  C-statistic decreased by 
0.3, 6.3, and 2.0 for He$\alpha$, Ly$\alpha$ and He$\beta$,
respectively, when compared with that shown in
table~\ref{tab:width_obs234}. Given these small improvements, we
conclude that it is difficult to distinguish the Voigt and Gaussian line
shapes using the present data.

After integrating the data of the entire SXS FOV ($60~{\rm kpc}\times60~{\rm kpc}$), no clear deviations from Gaussianity were found. This may be because deviations are spatially averaged and smeared out. To investigate the line profile in smaller areas, we extracted spectra from several $2\times2$ pixel ($20\times20~{\rm kpc}$) regions, and analyzed the Fe He$\alpha$ {\it w} profiles similarly. We found no common residuals clearly seen in Obs~2, 3, and 4 when the spectra of the pixels that corresponded to the same or similar sky regions were compared. We also separated the data into two groups, the central region (including the AGN) and the outer region, but obtained similar results. Finally, as independent indicators, the skewness and the kurtosis of the line profiles were calculated, and they were broadly consistent with those of Gaussian. No clear deviation from Gaussianity was found.

\subsection{Ion temperature measurements}
\label{sec:iontemperature}

In the analysis presented in previous sections, the observed line profiles are analyzed assuming that the ions are in thermal equilibrium with electrons and share the same temperature. High-resolution spectra by Hitomi provide us the first opportunity to directly test this assumption for galaxy clusters. As discussed in section~\ref{sec:discussion}, equilibration between electrons and ions takes longer than thermalization of the electron and ion distributions. A difference between the ion and electron temperatures may indicate a departure from thermal equilibrium.

The LOS velocity dispersion due to an isotropic thermal motion of ions is given by $\sigma_{\rm th} = \sqrt{k T_{\rm ion}/m_{\rm ion}}$, where $k$ is the Boltzmann constant, $T_{\rm ion}$ is the ion kinetic temperature, and $m_{\rm ion}$ is the ion mass. The LOS velocity dispersion from random hydrodynamic gas motions including turbulence, $\sigma_{\rm v}$, is assumed common for all the elements. Since only the former depends on $m_{\rm ion}$, one can in principle measure $\sigma_{\rm th}$ (i.e., $T_{\rm ion}$) and $\sigma_{\rm v}$ separately by combining the widths of lines originating from different heavy elements. For example, $kT_{\rm ion} = 4$~keV corresponds to $\sigma_{\rm th}=83, ~98, ~110, ~120$~km~s$^{-1}$ for Fe, Ca, S, and Si, respectively. These thermal velocities tend to be smaller than $\sigma_{\rm v}$ even for the lightest of currently observed elements, making the measurement of $T_{\rm ion}$ challenging. In what follows, we assume that the ions share a single kinetic temperature for simplicity.

\begin{figure*}[t]
\begin{minipage}{0.5\hsize}
 \begin{center}
  \includegraphics[width=8.5cm]{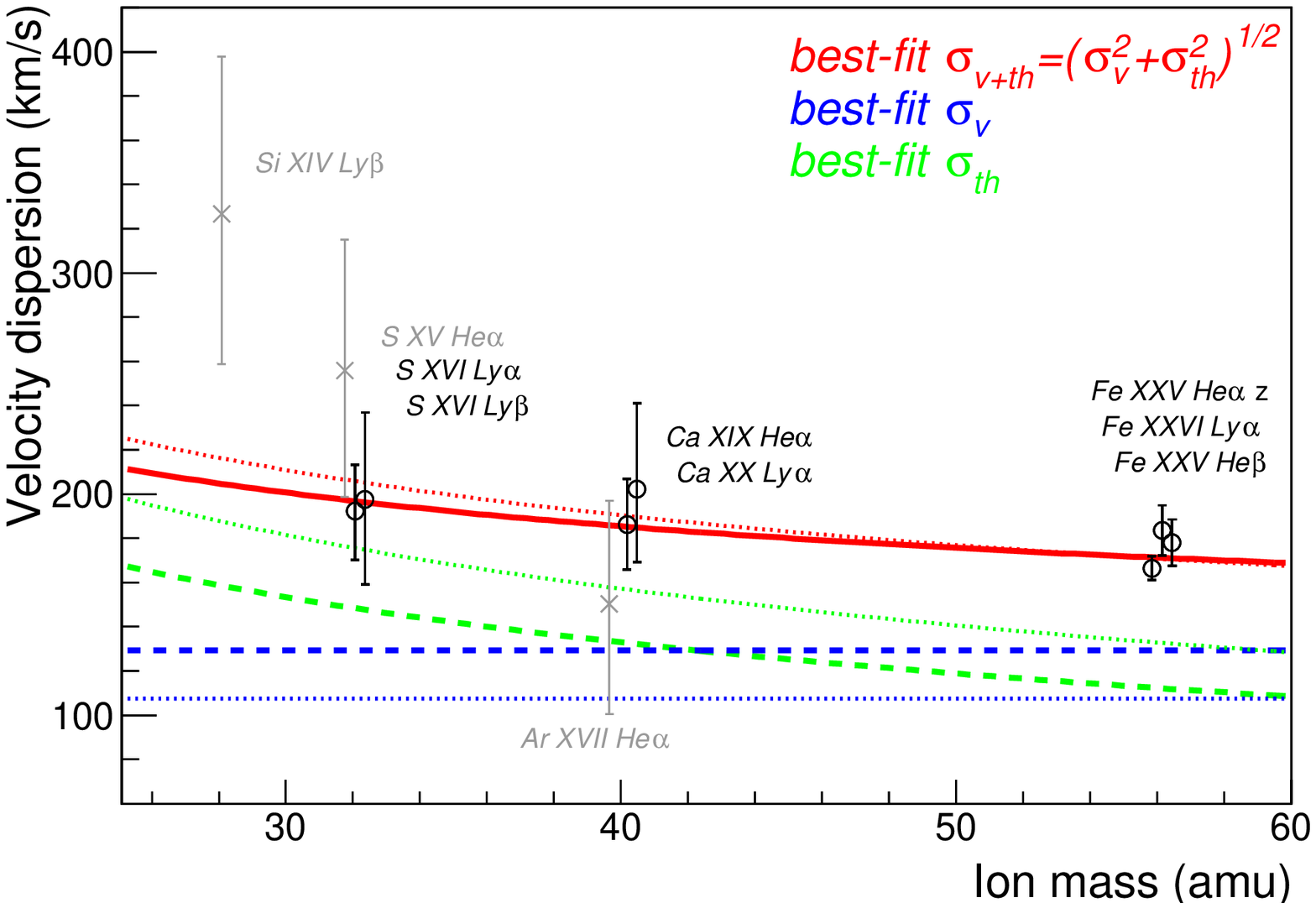}
 \end{center}
\end{minipage}
\begin{minipage}{0.5\hsize}
 \begin{center}
  \includegraphics[width=7cm]{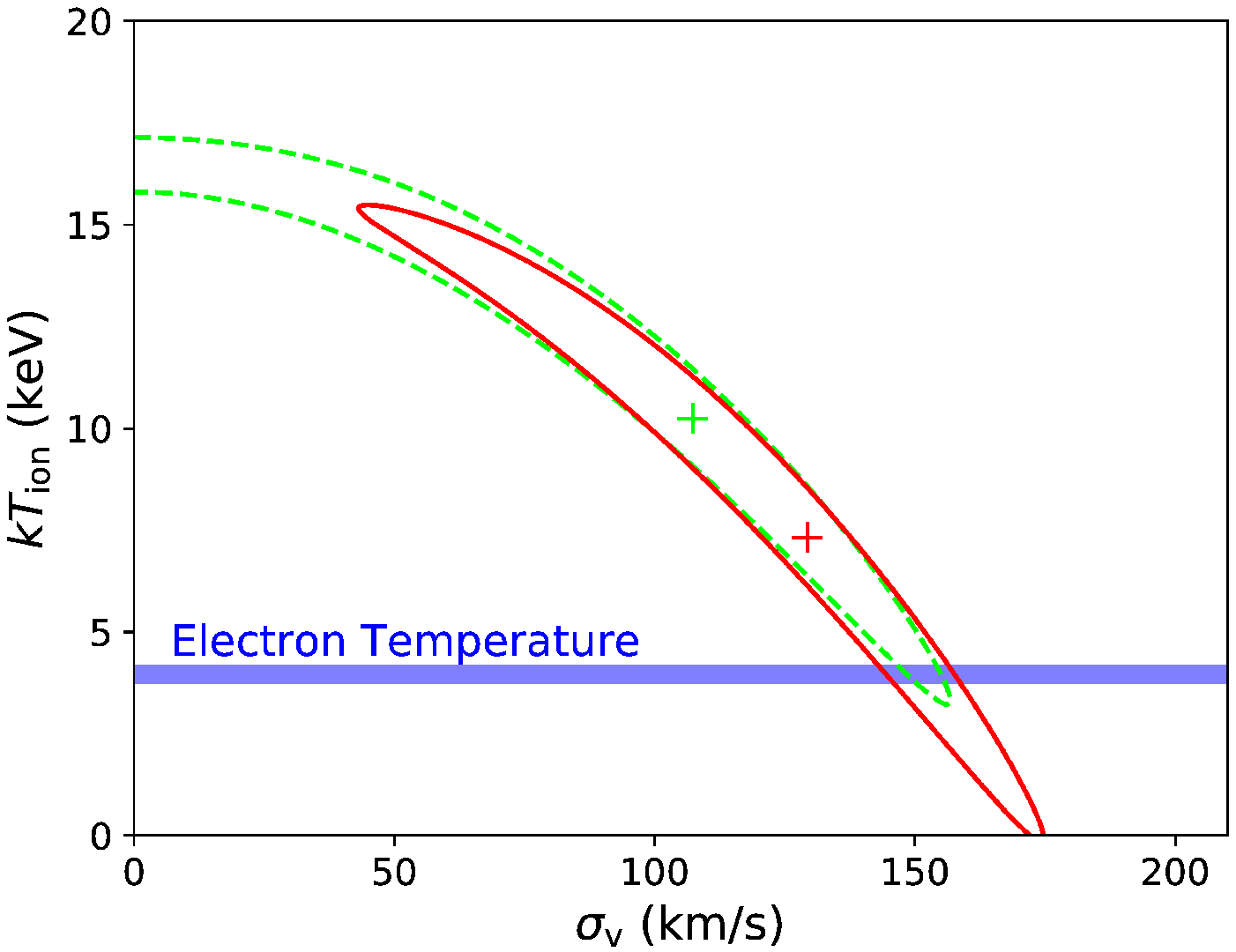}
 \end{center}
\end{minipage}
\caption{{\it Left:} The total velocity dispersion $\sigma_{\rm v+th}$ of bright lines as a function of the ion mass in the atomic mass unit (amu). For clarity, the data points for the same element are slightly shifted horizontally. Black circles and gray crosses denote the lines detected at more than $10\sigma$ significance and at $5-10 \sigma$ significance, respectively. Solid and dashed lines show the best-fit relation $\sigma_{\rm v+th}=(\sigma_{\rm th}^2 + \sigma_{\rm v}^2)^\frac{1}{2}$ (red solid) and its components $\sigma_{\rm th}$ (green dashed) and $\sigma_{\rm v}$ (blue dashed) for the $>10\sigma$ lines. Dotted lines are the best-fit relation $\sigma_{\rm v+th}$ (red dotted) and its components $\sigma_{\rm th}$ (green dotted) and $\sigma_{\rm v}$ (blue dotted) for the $>5\sigma$ lines. {\it Right:} The 68\% confidence regions of $kT_{\rm ion}$ and $\sigma_{\rm v}$ for two parameters of interest ($\Delta\chi^2=2.3$) with a plus marking the best-fit values. Red solid and green dashed contours represent the results for the $>10\sigma$ and $>5\sigma$ lines, respectively. For reference, the blue horizontal bar indicates the range of the electron temperature measured in T~paper.}
\label{fig:fit_Tion_empirical}
\end{figure*}

The left panel of figure~\ref{fig:fit_Tion_empirical} shows the total velocity dispersion $\sigma_{\rm v+th}$ of lines detected at more than $5\sigma$ significance listed in table~\ref{tab:width_obs234}. Unreliable measurements marked by notes 1--6 in table~\ref{tab:width_obs234} have been excluded. The lines from different elements show nearly consistent velocity dispersions with a weakly-decreasing trend with ion mass. They are fit by $\sigma_{\rm v+th}=(\sigma_{\rm v}^2 + \sigma_{\rm th}^2)^\frac{1}{2}$ varying $T_{\rm ion}$ and $\sigma_{\rm v}$ as free parameters. The best-fit values are $kT_{\rm ion} = 10.2^{+5.0}_{-4.6}$\,keV and $\sigma_{\rm v}=107^{+35}_{-58}$\,km~s$^{-1}$, with $\chi^2=7.104$ for 8 degrees of freedom. If only the most secure measurements at more than $10 \sigma$ significance (black circles in the left panel of figure~\ref{fig:fit_Tion_empirical}) are used, the best-fit values are $kT_{\rm ion} = 7.3^{+5.3}_{-5.0}$\,keV and $\sigma_{\rm v}=129^{+32}_{-45}$\,km~s$^{-1}$, with $\chi^2=2.640$ for 5 degrees of freedom. If we vary just a single parameter $\sigma_{\rm v}$ by setting $\sigma_{\rm th}=0$, we obtain $\sigma_{\rm v}=174.3^{+4.1}_{-4.2}$ km~s$^{-1}$ with $\chi^2=12.20$ for 9 degrees of freedom from the $> 5 \sigma$ lines, and $\sigma_{\rm v}=173.4\pm4.2$ km~s$^{-1}$ with $\chi^2=4.848$ for 6 degrees of freedom from the $> 10 \sigma$ lines.

The red solid and green dashed contours in the right panel of figure~\ref{fig:fit_Tion_empirical} show the 68\% confidence regions of $T_{\rm ion}$ and $\sigma_{\rm v}$ for the $> 10 \sigma$ lines and the $> 5 \sigma$ lines, respectively. As expected, a negative correlation is found between $T_{\rm ion}$ and $\sigma_{\rm v}$. Albeit with large errors, the inferred ion temperature is consistent within the 68\% confidence level with the electron temperature reported in T~paper. The calibrated SXS FWHM has a systematic error of $\sim$0.15~eV (see appendix~\ref{sec:systematic}), which does not alter the results of this subsection. The present errors are dominated by the uncertainties of the widths of the lines in the low energy (2--4~keV) band; higher significance data at lower energies and inclusion of lighter elements will be crucial for improving the measurement.

\begin{figure}[h]
\begin{center}
 \includegraphics[height=8cm,angle=-90]{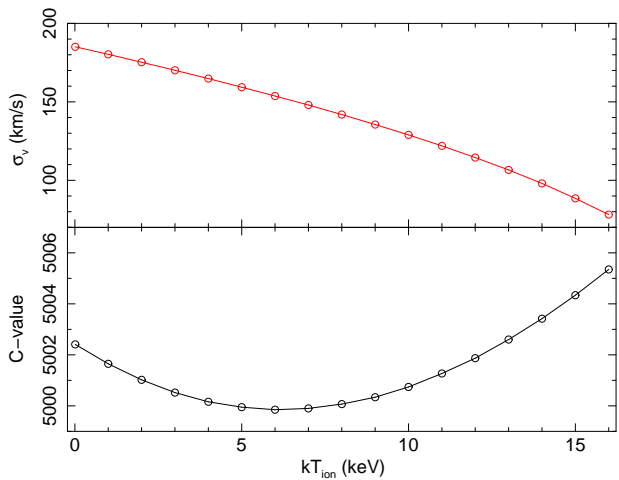}
\end{center}
\caption{Results of fitting the entire spectrum with a plasma code SPEX. Top and bottom panels show the optimal values of $\sigma_{\rm v}$ and C-statistic, respectively, for a given value of $T_{\rm ion}$.}
\label{fig:fit_Tion_spex}
\end{figure}

For comparison, we also infer the ion temperature by fitting the entire spectrum with a plasma code, SPEX v3.03.00 \citep{kaastra96}. Here we apply a gain correction using equation (A1) of Atomic~paper to match the observed line energies to those implemented in SPEX. We fit the spectrum with models of the collisional ionization equilibrium plasma, the central AGN, and the NXB components. For the central AGN, we adopt the model parameters determined by AGN~paper. We exclude the energy band covering the Fe XXV He$\alpha$ {\it w} line to eliminate the effect of resonance scattering. Figure \ref{fig:fit_Tion_spex} shows the optimal values of C-statistic and $\sigma_{\rm v}$ for a given value of $T_{\rm ion}$.  The best-fit values are $kT_{\rm ion}=6.0^{+4.2}_{-3.7}$\,keV and $\sigma_{\rm  v}=153^{+21}_{-27}$\,km~s$^{-1}$, with the C-statistic value of 4999.86 for 4653 degrees of freedom. Again a negative correlation between $T_{\rm ion}$ and $\sigma_{\rm v}$ is found. These results are consistent with those derived from a set of bright lines shown in figure \ref{fig:fit_Tion_empirical}.

Note that a similar analysis using SPEX is also performed in Atomic~paper. They present the results when the Fe XXV He$\alpha$ {\it w} line is included in the fit. Since this line is likely subject to resonance scattering (RS~paper), the fitted value of $T_{\rm ion}$ depends on how the radiative transfer effect is taken into account. They show that a simple absorption model implemented in SPEX yields the value of $T_{\rm ion}$ in good agreement with $T_{\rm e}$ (see section~7.1 of Atomic~paper for details).

\section{Discussion}
\label{sec:discussion}

\subsection{The origin of gas motions}
\label{sec:origin}

The Hitomi SXS observations provided the first direct measurements of the LOS velocities and velocity dispersions of the hot ICM in the core of the Perseus cluster. Using the optically thin emission lines, we find that the LOS velocity dispersion peaks toward the cluster center and around the prominent northwestern `ghost' bubble, reaching $\sigma_{\rm v}\sim200$~km~s$^{-1}$. These velocity dispersion peaks are seen in both PSF-corrected and uncorrected maps. Outside of these peaks, the LOS velocity dispersion appears constant at $\sigma_{\rm v}\sim100$~km~s$^{-1}$. Note that the velocity dispersion peak at the center is seen in the maps derived by both methods, excluding and including the resonance {\it w} line (appendix~\ref{sec:details}). The peak toward the ghost bubble is not seen when the {\it w} line is used for the velocity fits (appendix~\ref{sec:details}), so its existence is less certain.

The maximum velocity of 100~km~s$^{-1}$ determined from line shifts within the investigated area indicates that the velocity of large scale flows is at least $v_{\rm bulk}=100$~km~s$^{-1}$. While some theoretical arguments predict a velocity offset of the order of $\sim100$~km~s$^{-1}$ between the central galaxy and the ICM \citep{inoue14}, the zero point of our observed bulk shear is consistent with the redshift of NCG~1275. We note that as the photons produced within the central $r\sim100$~kpc climb up the gravitational potential well of the cluster, they are also affected by a gravitational redshift of $\sim 20$~km~s$^{-1}$. This shift should be considered in the absolute value of each redshift measurement. The $v_{\rm bulk}$ values are relative values between NGC~1275 and the ICM, and so the gravitational redshift is mostly canceled out. The relative gravitational redshift across the FOV is $\sim 5$~km~s$^{-1}$.

During the process of hierarchical structure formation, turbulent gas motions are driven on Mpc scales by mergers and accretion flows which convert their kinetic energy into turbulence \citep[e.g.][]{bruggen2015}. These turbulent motions then cascade down from the driving scales to dissipative scales, heating the plasma, (re-)accelerating cosmic-rays, and amplifying the magnetic fields \citep[e.g.][]{brunetti2007,miniati2015}. In the Perseus cluster, turbulence is also likely to contribute to powering the radio emission of the minihalo \citep{burns92,sijbring93,walker2017} by re-accelerating the relativistic electrons originating from the AGN and/or hadronic interactions \citep[e.g.][]{gitti2002,ZuHone13}.

Turbulence is also expected to be driven on smaller scales by the AGN, galaxy motions, gas sloshing, and hydrodynamic and magneto-thermal instabilities in the ICM \citep[e.g.][]{churazov2002,Gu13,mendygral2012,ichinohe17,ZuHone13,zuhone17}. The low, relatively uniform velocity dispersion observed in the Perseus core is also consistent with that expected for turbulence induced in the cool core by sloshing \citep{ZuHone13}. Several cold fronts are seen in the Perseus X-ray images \citep{Churazov03,simionescu2012,walker2017}, which reveal a sloshing core. If the observed velocity dispersion is indeed mostly sloshing induced, then an interesting prediction for future observations is that the observed dispersion will abruptly change across the cold fronts, which are mostly located outside the Hitomi FOV.

The observed peaks in $\sigma_{\rm v}$ appear to indicate that gas motions are driven both at the cluster center by the current AGN inflated bubbles and by the buoyantly rising ghost bubbles with diameters of $\sim25$~kpc. The observed peaks in $\sigma_{\rm v}$ could be due to superposed streaming motions around the bubbles and turbulence. This observation appears to contradict models in which gas motions are sourced only at the center (during the initial stages of bubble inflation) or only by structure formation. These results may indicate that both the current AGN inflated bubbles in the cluster center and the buoyantly rising ghost bubbles are driving gas motions in the Perseus cluster.

Part of the observed large scale motions of $v_{\rm bulk}\sim 100$~km~s$^{-1}$ might be due to streaming motions around and in the wakes of buoyantly rising bubbles as well. As already pointed out in H16, to the north of the core, the trend in the LOS velocities of the ICM is consistent with the trend in the velocities of the molecular gas within the northern optical emission line filaments \citep{salome2011}. These trends are consistent with the model where the optical emission line nebulae and the molecular gas result from thermally unstable cooling of low entropy gas uplifted by buoyantly rising bubbles \citep[e.g.][]{hatch2006,mcnamara2016}.

However, most of the bulk motions are likely driven by the gas sloshing in the core of the Perseus cluster \citep{Churazov03,walker2017,zuhone17}. The gas sloshing observed in the innermost cluster core, $r\lesssim100$~kpc, might be due to strong AGN outbursts \citep{Churazov03} or due to a disturbance of the cluster gravitational potential caused by a recent subcluster infall \citep[e.g.][]{markevitch2007} which is likely related to the large-scale sloshing in this system \citep{simionescu2012}. The molecular gas can be advected by the sloshing hot gas, resulting in their similar LOS velocities. The shearing motions associated with gas sloshing are also expected to contribute to the velocity dispersion observed throughout the investigated area.

Given the large density gradient in the core of the Perseus cluster, the effective length along the LOS from which the largest fraction of line flux (and measured line width) arises, $L_{\rm eff}$, is rapidly increasing as a function of radius. The increase of the effective length, $L_{\rm eff}$,  with growing projected distance $r$ implies that larger and larger eddies contribute to the observed line broadening. Therefore, as shown by \citet{zhuravleva2012}, for Kolmogorov-like turbulence driven on scales larger than $\sim 100$~kpc, we would expect to see a radially increasing LOS velocity dispersion. For example, for turbulence driven on scales of 200~kpc, we would expect a factor of 1.7 increase in the measured velocity dispersion over the radial range of 100~kpc (from the core out to $r\sim$100~kpc assuming the density profile of the Perseus cluster). The lack of observed radial increase of $\sigma_{\rm v}$ might indicate that the turbulence in the core of the Perseus cluster is driven primarily on scales smaller than $\sim 100$~kpc. The relative uniformity of the dispersion is also consistent with sloshing-induced turbulence, which is mostly limited to the cool core in the absence of large-scale disturbances such as a major merger \citep[see figures~14--16 in][]{ZuHone13}.

While turbulence on spatial scales $L < L_{\rm eff}$ will increase the observed line widths and the measured $\sigma_{\rm v}$, gas motions on scales $L > L_{\rm eff}$ will shift the line centroids. The superposition of large scale motions over the LOS within our extraction area should therefore lead to non-Gaussian features in the observed line shapes \citep[e.g.][]{Inogamov03}. The lack of evidence for non-Gaussian line shapes in the spectral lines extracted over a spatial scale of $\sim$100~kpc (see section~\ref{sec:nongaussianity}) indicates that the observed velocity dispersion is dominated by small scale motions and corroborates the conclusion that, in the core of the cluster, the driving scale of the turbulence is mostly smaller than $\sim100$~kpc.

From a suite of cosmological cluster simulations by \citet{nelson2014}, and an isolated high-resolution cluster simulation with cooling and AGN feedback physics by \citet{gaspari2012}, \citet{lau2017} generated a set of mock Hitomi SXS spectra to study the distribution and the characteristics of the observed velocities. They concluded that infall of subclusters and mechanical AGN feedback are the key complementary drivers of the observed gas motions. While the gentle, self-regulated mechanical AGN feedback sustains significant velocity dispersions in the inner innermost cool core, the large-scale velocity shear at $\gtrsim 50$~kpc is due to mergers with infalling groups. The comparison with their simulations also suggests that the AGN feedback is ``gentle'', with many small outbursts instead of a few isolated powerful ones \citep[see also][]{Fabian06,Fabian12,mcnamara2012,mcnamara2016}.  Similar conclusions were reached in the simulations by \citet{bourne2017}.

\subsection{Kinetic pressure support}
\label{sec:pressure}

One of the key implications of the gas velocities measured in section \ref{sec:analysis} is that hydrostatic equilibrium holds to better than 10\% near the center of the Perseus cluster. The results presented in figure~\ref{fig:velocity_psfcor} suggest that, if the observed velocity dispersion is due to isotropic turbulence, the inferred range of $\sigma_{\rm v} \sim$100--200~km~s$^{-1}$ corresponds to 2--6\% of the thermal pressure support of the gas with $kT = 4$~keV.

The large scale bulk motion will also contribute to the total kinetic energy. Assuming further that the observed line shifts are due to bulk motions with velocities of $v_{\rm bulk} = 100$~km~s$^{-1}$ with respect to the cluster center, the fraction of the kinetic to thermal energy density is
\begin{eqnarray}
\frac{\epsilon_{\mathrm{kin}}}{\epsilon_{\mathrm{therm}}}
= \frac{\mu m_{\rm p}(3\sigma^{2}_{\rm v}+v^{2}_{\rm bulk})}{3kT}
\sim 0.02 - 0.07,\label{eq:kintotherm}
\end{eqnarray}
for $kT=4$~keV, where $\mu=0.6$ is the mean molecular weight, and $m_{\rm p}$ is the proton mass. The expression can also be rewritten as $\epsilon_{\mathrm{kin}}/\epsilon_{\mathrm{therm}}=(\gamma/3)\mathcal{M}^2$, where the Mach number $\mathcal{M}=v_{\rm 3D~eff}/c_{\rm s} \sim 0.19-0.35$, $v_{\rm 3D~eff}=\sqrt{3\sigma^{2}_{\rm v}+v^{2}_{\rm bulk}}$ is the effective three dimensional velocity, $c_{\mathrm{s}}=\sqrt{\gamma kT/\mu m_{\mathrm{p}}} = 1030 (kT/4~{\rm keV})^{1/2}$~km~s$^{-1}$ is the sound speed, and $\gamma=5/3$ is the adiabatic index. The small amount of the kinetic energy density supports the validity of total cluster mass measurements under the assumption of hydrostatic equilibrium \citep[e.g.,][]{Allen11}, at least in the cores of galaxy clusters.

We note, however, that if the velocity dispersion is mostly sloshing induced, we might be underestimating the kinetic energy density. Sloshing in the Perseus cluster appears to be mostly in the plane of the sky and \citet{ZuHone13} show that such a relative geometry results in a total kinetic energy being a factor (5--6)$\sigma_\mathrm{v}^2$, compared to the factor of 3 in Equation~\ref{eq:kintotherm} for isotropic motions. This would change the upper bound of the kinetic to thermal pressure ratio to 0.11--0.13.

\subsection{Maintaining the balance between cooling and heating}
\label{sec:dissip}

The gas in the core of galaxy clusters appears to be in an approximate global thermal balance, which is likely maintained by several heating and energy transport mechanisms taking place simultaneously. One possible source of heat is the central AGN. Relativistic jets, produced by the central AGN drive weak shocks with Mach numbers of 1.2--1.5 \citep[e.g.][]{forman2005,forman2007,forman2017,nulsen2005,simionescu2009a,million2010,randall2011,randall2015} and inflate bubbles of relativistic plasma in the surrounding X-ray-emitting gas \citep[e.g.][]{Boehringer93,Churazov00,Fabian03,Fabian06,birzan2004,dunn2005,forman2005,forman2007,dunn2006,dunn2008,rafferty2006,McNamara07}. The bubbles appear to be inflated gently, with most of the  energy injected by the AGN going into the enthalpy of bubbles and only $\lesssim20$\% carried by shocks \citep{forman2017,zhuravleva2016,tang2017}. After detaching from the jets, the bubbles rise buoyantly and they often entrain and uplift large quantities of low entropy gas from the innermost regions of their host galaxies \citep{simionescu2008,simionescu2009b,kirkpatrick2009,kirkpatrick2011,Werner10,werner2011,mcnamara2016}. All of this activity is believed to take place in a tight feedback loop, where the hot ICM cools and accretes onto the central AGN, leading to the formation of jets which heat the surrounding gas, lowering the accretion rate, reducing the feedback, until the accretion eventually builds up again \citep[for a review see][]{McNamara07,Fabian12}.

Many questions regarding the energy transport from the bubbles to the ICM remain. Part of the energy might be transported by turbulence generated in situ by bubble-driven gravity waves oscillating within the gas \citep[e.g.][]{churazov01}. While g-modes are efficient at spreading the energy azimuthally, they are not able to transport energy radially \citep[e.g.][]{reynolds2015}. Energy can also be carried by bubble-generated sound waves \citep{Fabian03,Fujita05,Sanders07}, which could propagate fast enough to heat the core \citep{fabian2017}. The energy from bubbles can also be transported to the ICM by cosmic ray streaming and mixing \citep[e.g.][]{loewenstein1991,guo2008,fujita11,pfrommer2013,ruszkowski2017,jacob2017} or by mixing of the bubbles \citep[e.g.][]{hillel2016,hillel2017}.
 
The Hitomi SXS observation of the Perseus cluster allows us to explore the role of the dissipation of gas motions in keeping the ICM from cooling. As discussed in section~\ref{sec:origin}, substantial part of the kinetic energy density in the core of the Perseus cluster could be generated by the AGN, which appears to produce a peak in $\sigma_{\rm v}$ toward the cluster center and possibly around the prominent northwestern ghost bubble. Heating by dissipation of turbulence, induced by buoyantly rising (at a significant fraction of the sound speed) AGN-inflated bubbles, provides an attractive regulating mechanism for balancing the cooling of the ICM through a feedback loop \citep[e.g][]{McNamara07}. The rising bubbles are expected to generate turbulence in their wakes and excite internal waves, which propagate efficiently in azimuthal directions and decay to volume-filling turbulence. Based on the analysis of surface brightness fluctuations measured with Chandra, \citet{Zhuravleva14} showed that the heating rate from the dissipation of gas motions is capable of balancing the radiative cooling at each radius in the Perseus cluster. The direct measurements of the velocity dispersion by the Hitomi SXS are broadly consistent with these previous indirect deductions \citep[see figure~11 in][which compares the Chandra results with the earlier measurements reported by H16]{zhuravleva2017}. Note, however, that the dissipation of observed gas motions is capable of balancing radiative cooling only if (i) these motions dissipate in less than 10\% of the cooling timescale ($\sim$ Gyr) and (ii) they are continuously replenished over the age of the Perseus cluster.

Numerical simulations by \citet{ZuHone2010} showed that gas sloshing can facilitate the heat inflow into the core from the outer, hotter cluster gas via mixing, which can be enough to offset radiative cooling in the bulk of the cool core, except the very center. While the dissipation of turbulence induced by mergers \citep{Fujita04} or galaxy motions \citep{balbus1990,Gu13} could also contribute to heating the ICM, they would be unable to maintain a fine-tuned feedback loop.

\subsection{Thermal equilibrium between electrons and ions}

We performed the first measurement of the ICM ion temperature, based on the thermal broadening of the emission lines. We find the ion temperature to be consistent with the electron temperature, albeit with large uncertainties. Equilibration via Coulomb collisions between the ions and electrons takes place over the timescale given by
\begin{eqnarray}
t_{\rm eq} \sim  6 \times10^6 \,{\rm yr} \left(\frac{n_{\rm e}}{10^{-2}\,{\rm cm}^{-3}}\right)^{-1}\left(\frac{kT}{4 ~{\rm keV}}\right)^{3/2},
\label{eq-tie}
\end{eqnarray}
where $n_{\rm e}$ is the number density of electrons \citep{spitzer65,zeldovich66}. The equilibration time scales for electrons and for the ions are much shorter by factors of about $m_{\rm p}/m_{\rm e} \simeq 1800$ and $\sqrt{m_{\rm p}/m_{\rm e}} \simeq 43$, respectively, where $m_{\rm p}$ is the proton mass and $m_{\rm e}$ is the electron mass. Because the ions in the ICM are almost fully ionised and the rate of Coulomb collisions is proportional to the electric charge squared, their equilibration time scale is governed by that of protons; the ions equilibrate with protons faster than protons among themselves. Therefore, if the ICM has equilibrated via Coulomb collisions, equation (\ref{eq-tie}) gives a lower limit to the time elapsed since the last major heat injection. This timescale is much shorter than any relevant merger or AGN-related timescales, thus we did not expect to find a discrepancy between $T_\mathrm{e}$ and $T_\mathrm{ion}$.


\section{Conclusions}
\label{sec:conclusions}

In this paper, we have presented Hitomi observations of the atmospheric gas motions in the core, $r\lesssim100$~kpc, of the Perseus galaxy cluster. Our findings are summarized as follows.

\begin{enumerate}
 \item We have resolved and measured the line widths of He-like and H-like ions of Si, S, Ar, Ca, and Fe in the hot ICM for the first time.

 \item Using the optically thin emission lines and after correcting for the point spread function of the telescope, we find that the line-of-sight velocity dispersion of the hot gas is mostly low and uniform. The line-of-sight velocity dispersion of the hot gas reaches maxima of approximately 200~km~s$^{-1}$ toward the central AGN and toward the AGN inflated north-western `ghost' bubble. Elsewhere within the observed region, the velocity dispersion appears nearly uniform at $\sigma_{\rm v} \sim 100$~km~s$^{-1}$. The systematic uncertainty affecting the best-fit line-of-sight velocity dispersion values is $\lesssim$20~km~s$^{-1}$ (gain), $\lesssim$3~km~s$^{-1}$ (line spread function) and $\lesssim$5~km~s$^{-1}$ (PSF shape) in most cases.

 \item We detect a large scale bulk velocity gradient with an amplitude of $\sim 100$~km~s$^{-1}$ across the cluster center, consistent with sloshing induced motions.

 \item The mean redshift of the hot atmosphere is consistent with that of the stars of the central galaxy NGC~1275.

 \item The shapes of well-resolved optically thin emission lines are consistent with Gaussian. The lack of evidence for non-Gaussian line shapes indicates that the observed velocity dispersion is dominated by small scale motions. Our results imply that the driving scale of turbulence is mostly smaller than $\sim100$~kpc.

 \item If the observed gas motions are isotropic, the kinetic pressure support in the cluster core is smaller than 10\% of the thermal pressure.

 \item Combining the widths of the lines formed from various elements, we have obtained the first direct constraints on the thermal motions of the ions in the hot ICM. We find no evidence of deviation between the ion temperature and the electron temperature.

\end{enumerate}

Owing to the short lifetime of Hitomi, our results are restricted to the central region of a single galaxy cluster. Future X-ray calorimeter missions, e.g., the X-ray Astronomy Recovery Mission (XARM) and Athena \citep{nandra13}, will be crucial for extending the measurements to larger radii and a larger number of clusters, thereby providing further insights into the dynamics of galaxy clusters.

\section*{Author Contributions}
Y. Ichinohe and S. Ueda led this study and wrote the final manuscript along with T. Kitayama, B. McNamara, N. Werner, R. Fujimoto, S. Inoue, M. Markevitch, and C. Kilbourne.
Y. Ichinohe and S. Ueda performed the analysis of section~\ref{sec:velocity} and appendices~\ref{sec:systematic}, \ref{sec:xcm}, \ref{sec:arfs}, and \ref{sec:heb}.
R. Fujimoto and K. Tanaka conducted the analysis of sections~\ref{sec:spectra} and \ref{sec:nongaussianity}.
S. Inoue and T. Kitayama performed the analysis of section~\ref{sec:iontemperature}.
N. Werner, B. McNamara, and I. Zhuravleva provided various inputs to section \ref{sec:discussion}.
R. Canning measured the new redshift of the central galaxy NGC~1275 presented in appendix \ref{sec:redshift}.
M. Markevitch performed the analysis of appendix~\ref{sec:maxim}.
Q. Wang contributed to the analysis of appendix~\ref{sec:xcm}.
T. Tamura, N. Ota, M. Tsujimoto, K. Sato, and S. Nakashima contributed to the velocity mapping analysis and studies on systematic uncertainties.
R. Fujimoto, C. Kilbourne, and S. Porter achieved the development, integration tests, and in-orbit operation of the SXS.
Y. Maeda supported the evaluation of the PSF scattering effect.
T. Hayashi, S. Kitamoto, and I. Zhuravleva evaluated the impact of the gravitational redshift.
The science goals of Hitomi were discussed and developed over more than 10 years by the ASTRO-H Science Working Group (SWG), all members of which are authors of this manuscript. All the instruments were prepared by joint efforts of the team. The manuscript was subject to an internal collaboration-wide review process. All authors reviewed and approved the final version of the manuscript.

\begin{ack}
We are grateful to the anonymous referee for helpful suggestions and comments.
We acknowledge Yuya Kinoshita for his detailed analysis on the non-Gaussianity in 2x2 pixel scale and evaluation of skewness and kurtosis, Yu Kai, Ayumi Tsuji, and Tomohiro Nakano for supporting data analysis.
We thank the support from the JSPS Core-to-Core Program.
We acknowledge all the JAXA members who have contributed to the ASTRO-H (Hitomi)
project.
All U.S. members gratefully acknowledge support through the NASA Science Mission
Directorate. Stanford and SLAC members acknowledge support via DoE contract to SLAC
National Accelerator Laboratory DE-AC3-76SF00515. Part of this work was performed under
the auspices of the U.S. DoE by LLNL under Contract DE-AC52-07NA27344.
Support from the European Space Agency is gratefully acknowledged.
French members acknowledge support from CNES, the Centre National d'\'{E}tudes Spatiales.
SRON is supported by NWO, the Netherlands Organization for Scientific Research.  Swiss
team acknowledges support of the Swiss Secretariat for Education, Research and
Innovation (SERI).
The Canadian Space Agency is acknowledged for the support of Canadian members.  
We acknowledge support from JSPS/MEXT KAKENHI grant numbers 15J02737,
15H00773, 15H00785, 15H02090, 15H03639, 15H05438, 15K05107, 15K17610,
15K17657, 16J00548, 16J02333, 16H00949, 16H06342, 16K05295, 16K05296,
16K05300, 16K13787, 16K17672, 16K17673, 17J07948, 21659292, 23340055, 23340071,
23540280, 24105007, 24244014, 24540232, 25105516, 25109004, 25247028,
25287042, 25400236, 25800119, 26109506, 26220703, 26400228, 26610047,
26800102, JP15H02070, JP15H03641, JP15H03642, JP15H06896,
JP16H03983, JP15J01845, JP16K05296, JP16K05309, JP16K17667, and JP16K05296.
The following NASA grants are acknowledged: NNX15AC76G, NNX15AE16G, NNX15AK71G,
NNX15AU54G, NNX15AW94G, and NNG15PP48P to Eureka Scientific.
H. Akamatsu acknowledges support of NWO via Veni grant.  
C. Done acknowledges STFC funding under grant ST/L00075X/1.  
A. Fabian and C. Pinto acknowledge ERC Advanced Grant 340442.
P. Gandhi acknowledges JAXA International Top Young Fellowship and UK Science and
Technology Funding Council (STFC) grant ST/J003697/2. 
Y. Ichinohe, K. Nobukawa, H. Seta, S. Inoue, and T. Hayashi are supported by the Research Fellow of JSPS for Young
Scientists.
N. Kawai is supported by the Grant-in-Aid for Scientific Research on Innovative Areas
``New Developments in Astrophysics Through Multi-Messenger Observations of Gravitational
Wave Sources''.
S. Kitamoto is partially supported by the MEXT Supported Program for the Strategic
Research Foundation at Private Universities, 2014-2018.
B. McNamara and S. Safi-Harb acknowledge support from NSERC.
T. Dotani, T. Takahashi, T. Tamagawa, M. Tsujimoto and Y. Uchiyama acknowledge support
from the Grant-in-Aid for Scientific Research on Innovative Areas ``Nuclear Matter in
Neutron Stars Investigated by Experiments and Astronomical Observations''.
Q. Wang was supported by Chandra grants GO3-14144Z, GO5-16147Z and AR5-16013X.
N. Werner is supported by the Lend\"ulet LP2016-11 grant from the Hungarian Academy of
Sciences.
D. Wilkins is supported by NASA through Einstein Fellowship grant number PF6-170160,
awarded by the Chandra X-ray Center, operated by the Smithsonian Astrophysical
Observatory for NASA under contract NAS8-03060.

We thank contributions by many companies, including in particular, NEC, Mitsubishi Heavy
Industries, Sumitomo Heavy Industries, and Japan Aviation Electronics Industry. We acknowledge
Google Inc. for their web-based services which really boosted our productivity. Finally,
we acknowledge strong support from the following engineers.  JAXA/ISAS: Chris Baluta,
Nobutaka Bando, Atsushi Harayama, Kazuyuki Hirose, Kosei Ishimura, Naoko Iwata, Taro
Kawano, Shigeo Kawasaki, Kenji Minesugi, Chikara Natsukari, Hiroyuki Ogawa, Mina Ogawa,
Masayuki Ohta, Tsuyoshi Okazaki, Shin-ichiro Sakai, Yasuko Shibano, Maki Shida, Takanobu
Shimada, Atsushi Wada, Takahiro Yamada; JAXA/TKSC: Atsushi Okamoto, Yoichi Sato, Keisuke
Shinozaki, Hiroyuki Sugita; Chubu U: Yoshiharu Namba; Ehime U: Keiji Ogi; Kochi U of
Technology: Tatsuro Kosaka; Miyazaki U: Yusuke Nishioka; Nagoya U: Housei Nagano;
NASA/GSFC: Thomas Bialas, Kevin Boyce, Edgar Canavan, Michael DiPirro, Mark Kimball,
Candace Masters, Daniel Mcguinness, Joseph Miko, Theodore Muench, James Pontius, Peter
Shirron, Cynthia Simmons, Gary Sneiderman, Tomomi Watanabe; ADNET Systems: Michael
Witthoeft, Kristin Rutkowski, Robert S. Hill, Joseph Eggen; Wyle Information Systems:
Andrew Sargent, Michael Dutka; Noqsi Aerospace Ltd: John Doty; Stanford U/KIPAC: Makoto
Asai, Kirk Gilmore; ESA (Netherlands): Chris Jewell; SRON: Daniel Haas, Martin Frericks,
Philippe Laubert, Paul Lowes; U of Geneva: Philipp Azzarello; CSA: Alex Koujelev, Franco
Moroso.

\end{ack}

\bibliographystyle{aasjournal}
\bibliography{ref}

\appendix
\section{New redshift measurement of NGC~1275 using absorption lines}
\label{sec:redshift}

\begin{figure*}
\begin{center}
\includegraphics[width=8cm]{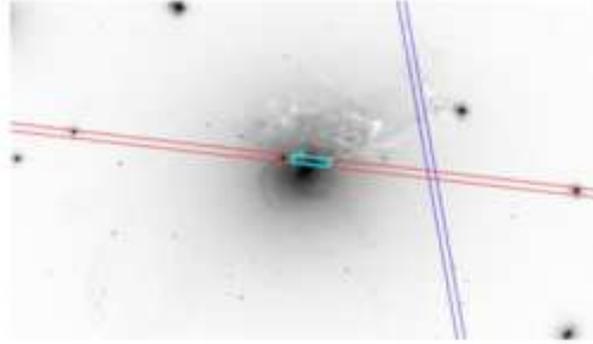}
\end{center}
\caption{The slit placement on NGC~1275. Our spectral extraction region is indicated with the cyan box on the red slit. Only the brighter parts of the low velocity system are extracted to avoid contamination by both a bright star and the high velocity system.}
\label{fig:ApA1}
\end{figure*}

Long slit spectroscopy was performed using the Intermediate dispersion Spectrograph and Imaging System (ISIS) at the 4.2~m William Herschel Telescope on the island of La Palma on 2007 December 29. The data were reduced using tailored IDL routines (adapted from the KRISIS IDL scripts by J.R. Mullaney 2008) for standard bias, flat field correction and wavelength calibration. The spectra were then traced and extracted separately on each frame using Gaussian and Lorentz profile fits in the cross-dispersion direction. Only the brighter parts of the low velocity system are extracted to avoid contamination by both a bright star which is in the slit and the high velocity system (see figure~\ref{fig:ApA1}). The spectra are median-combined. The wavelength calibration was checked and refined using bright sky Hg lines at air wavelengths of 4046.565\AA\ and 4358.335\AA. These features, especially at 4358\AA, are strong in our spectra and allow a finer, more precise wavelength calibration.

We fit the median combined R300B arm spectra using pPXF, which is an IDL program to extract the stellar kinematics or stellar population from absorption-line spectra of galaxies using the Penalized Pixel-Fitting method \citep[pPXF;][]{Cappellari04,cappellari17}. We fit Miles stellar population synthesis models with an IMF slope of 1.3 and metallicity values ranging from $-2.32$ to $+0.22$. The stellar kinematics is fit with the emission lines masked out. We obtain a best fit redshift of $z=0.017284\pm0.000039$ with only the statistical fitting uncertainties included. Including the upper and lower limits on wavelength calibration, we obtain $z=0.017284\pm0.00005$. For comparison, fitting the [O II] emission line doublet in the same region as the absorption lines gives $z=0.01697\pm0.00003$.

\section{Systematic uncertainty}
\label{sec:systematic}

\subsection{Gain uncertainty}
\label{sec:gain}

We achieved the systematic gain difference between Obs~3 and Obs~4 of $\lesssim0.3$~eV at 6.586~keV (the line centroid of Fe He$\alpha$ {\it w} in observer frame) with the standard pipeline gain correction processes alone. As the pointings of Obs~1 and Obs~2 were performed before the temperature of the helium tank reached near thermal equilibrium, an additional energy scale adjustment (\verb+sxsperseus+\footnote{https://heasarc.gsfc.nasa.gov/docs/hitomi/analysis/ahhelp/sxsperseus.html}), in addition to the standard pipeline gain correction, was applied to these datasets. As the FOV of Obs~2 overlaps with those of Obs~3 or Obs~4, we are able to compare the gain among these observations directly. After the gain adjustment, the data of Obs~2 have a gain offset of $\sim$2~eV at 6.586~keV, compared to Obs~3 (and Obs~4). As the FOV of Obs~1 does not overlap with those of Obs~2, 3 or 4, the absolute gain scale of Obs~1 is difficult to estimate. Considering the $\sim$2~eV gain offset of Obs~2, we think that the systematic uncertainty of the energy scale of Obs~1, is at least $\sim$2~eV relative to Obs~3. The pixel-to-pixel relative gain uncertainty within each single pointing is $\sim$0.5~eV. More details are described in \citet[][]{eckart17}.

\subsubsection{Effect of the gain uncertainty}
\label{sec:effect}

We investigated the effect of the gain uncertainty described in section~\ref{sec:gain} on the velocity measurements. We manually shifted the gain\footnote{We used \texttt{rmodel gain} command available in XSPEC, with \texttt{slope} $=1$ and \texttt{intercept} $=\Delta E$ where $\Delta E$ is the gain shift. We used the energy range of 6.4--6.7~keV.} and followed the same velocity fitting described in section~\ref{sec:velocity} to see how the result changes by the systematic gain difference. We shifted (1) the gain of all the Obs~1 data by $\pm$2~eV to account for the uncertainty of Obs~1 gain relative to Obs~3 and Obs~4 gain. (2) the gain of Reg~5 Obs~1 by $\pm$0.5~eV and at the same time the gain of Reg~6 Obs~1 by $\mp$0.5~eV for the pixel-to-pixel gain uncertainties within Obs~1. (3) the gain of Reg~0 Obs~3, Reg~0 Obs~4, Reg~1 Obs~3, Reg~1 Obs~4, Reg~2 Obs~3, Reg~2 Obs~4, Reg~3 Obs~3, Reg~3 Obs~4, Reg~4 Obs~3, and Reg~4 Obs~4 by $0.5~\mathrm{eV}/\sqrt{n}$, where $n$ is the number of pixels of each single region, twenty times with random signs in each trial, for relative gain uncertainties within Obs~3 and Obs~4.

\begin{figure*}
 \begin{center}
 \includegraphics[width=15cm]{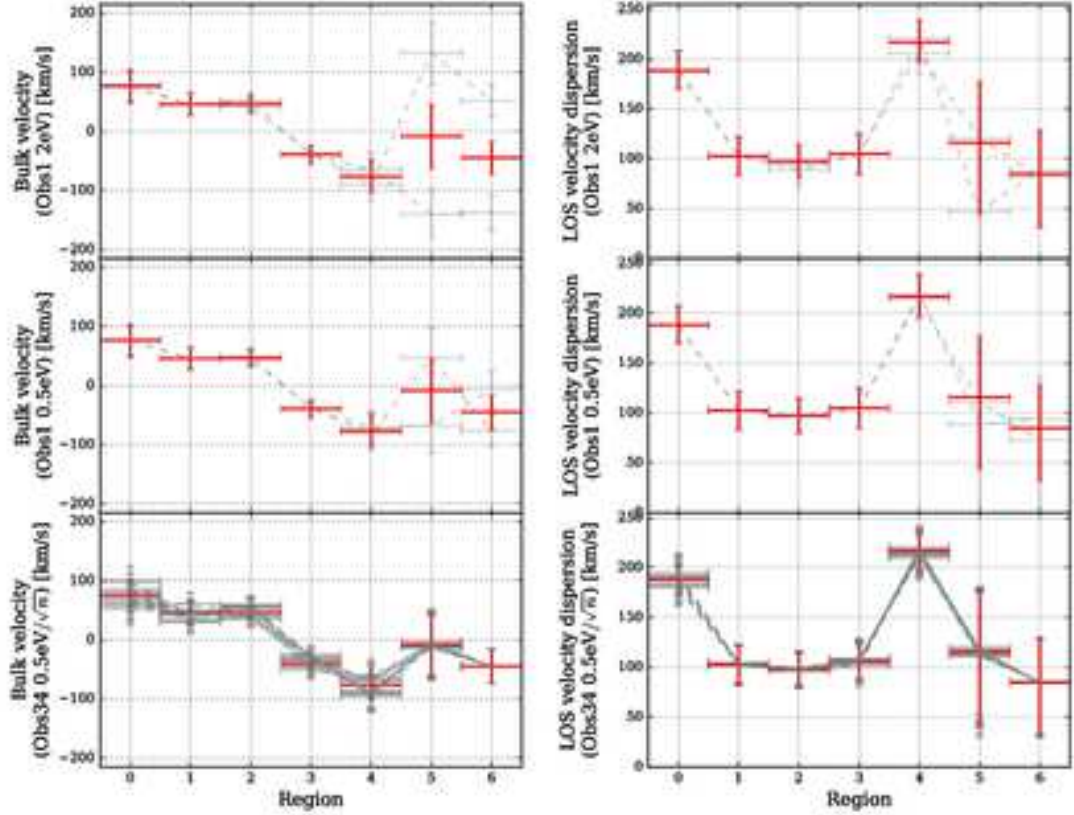} 
 \end{center}
 \caption{The best-fitting bulk velocities and LOS velocity dispersions after manually shifting the energy gain. {\it Top:} the effect of the uncertainty of Obs~1 gain relative to Obs~3 and Obs~4. {\it Middle:} the effect of the pixel-to-pixel gain uncertainty within Obs~1. {\it Bottom:} the effect of the pixel-to-pixel gain uncertainties within Obs~3 and Obs~4. The red crosses are the best-fitting values shown in table~\ref{tab:velocity}, and the grey crosses and dashed lines represent the best-fitting values after the gain shifts.}
\label{fig:plotgain}
\end{figure*}

The best-fitting bulk velocities and LOS velocity dispersions after the above mentioned gain shifts are shown in figure~\ref{fig:plotgain}. We found in every case that the LOS velocity dispersion does not change significantly from the nominal value ($\lesssim$20~km~s$^{-1}$ except for one case in Reg~5), although the best-fitting bulk velocity changes in proportion to the gain shifts.

\subsection{Effect of the line spread function uncertainty}
\label{sec:lsf}

We examined the uncertainty of line spread function (LSF) of the SXS and its effect on the measurement of LOS velocity dispersion. Due to the incomplete state of the SXS calibration at the time of these observations, it is not possible to determine a robust estimate of the uncertainty on the instrumental broadening. In H16, we conservatively estimated the range of possible FWHM values as 5$\pm$0.5~eV and set that as the 90\% confidence level. This estimate was based on variation in the calibration pixel LSF over time, how the array resolution compared with the calibration-pixel resolution during the later calibration measurement, and the difference in apparent line widths between Obs~2 and Obs~3. Even this conservative value corresponded to a smaller uncertainty on the velocity broadening at the Fe He-alpha lines than that due to the statistical uncertainty. For the current paper, we would like to be able to use a less conservative value, and to assess the impact of both estimates on our results. For the more optimistic estimate, we have chosen $\pm$0.15~eV, based on the dispersion of the resolutions of the individual pixels across the array during the later in-orbit calibration with 55Fe, and the premise that this dispersion represents pixel-dependent temporal variation more than intrinsic differences in the resolution.

The effect of its uncertainty on LOS velocity dispersion is expressed by
\begin{eqnarray}
\Delta \sigma_{\rm v} &\simeq& 3~{\rm km~s}^{-1}
\left(\frac{\sigma_{\rm v}}{100~{\rm km~s}^{-1}} \right)^{-1}
\left(\frac{W_{\rm inst}}{5~{\rm eV}} \right)
\left(\frac{\Delta W_{\rm inst}}{0.15~{\rm eV}}\right)
\left(\frac{E_{\rm obs}}{6.7~{\rm keV}} \right)^{-2},
\label{eq-deltavt2}
\end{eqnarray}
where $W_{\rm inst}$ is the FWHM of instrumental broadening and $\Delta W_{\rm inst}$ is an uncertainty of instrumental broadening in FWHM, assuming $\Delta W_{\rm inst} \ll W_{\rm inst}$ \citep[more details are shown in appendix of][]{kitayama-wp}. The effect is negligible.

\subsection{Effect of the PSF shape uncertainty}

We examined systematic uncertainties of LOS velocity dispersion introduced by the uncertainty of the PSF shape. As indicated in table~\ref{tab:psf}, the cross-term contribution from Sky~0 to Reg~1 Obs~3 is the largest among cross-term contributions. We found that the difference is typically $\lesssim$5~km~s$^{-1}$, except for Reg~1 ($\sim$10~km~s$^{-1}$) even when this cross-term was changed by $\pm30$\%, which is the maximum calibration uncertainty of the off-axis PSF normalizations between in the ground and in orbit \citep[][]{maeda17}. We also checked the effect of PSF uncertainty on the results of Obs~1 (Reg~5 and Reg~6). By changing the contribution from Sky~2 by $\pm 30$\% where is the largest contribution among the sky regions, we found that the difference is $\sim$5~km~s$^{-1}$, except for Reg~5 ($\sim$20~km~s$^{-1}$).

\subsection{Effect of the modeling uncertainty}
\label{sec:modeling}

We investigated the systematic uncertainty originating from plasma emission modeling. We examined the change of the best-fit redshift by fitting only Fe He$\alpha$ {\it w} line, which is not used in the velocity fitting in section~\ref{sec:velocity}. The analysis details are shown in appendix~\ref{sec:maxim}. This line has the highest counts among the He$\alpha$ complex. While the shape of this line can be affected by resonance scattering, the line centroid position is expected to be nearly unchanged. We obtained that the PSF-uncorrected bulk velocity of each regions are consistent between two methods except for in Reg~3 and Reg~6 (see table~\ref{tab:velocity_modelmix}). The offset of bulk velocity in these two regions is $\lesssim45$~km~s$^{-1}$. However, in Reg~6, when we modeled the {\it w} line using \verb+bapec+, we obtained a consistent bulk velocity with that shown in table~\ref{tab:velocity}. This suggests that the discrepancy originates from the emission modelling uncertainties. The effect of the modeling uncertainty on the bulk velocity measurements is therefore $\lesssim45$~km~s$^{-1}$.

\section{Details of velocity mapping}
\label{sec:details}

\subsection{Accounting for PSF scattering}
\label{sec:xcm}

We now describe how we accounted for PSF scattering in section~\ref{sec:velocity} in further detail. In the presence of steep X-ray surface brightness gradients, such as those in the cluster cool cores, the X-ray mirror PSF with a sharp core and broad wings \citep{okajima16} can cause significant flux contamination from the bright cluster peak into the lower-brightness regions at distances much greater than the nominal HPD of 1.2~arcmin. It is therefore essential to take PSF scattering into account even if the regions of interest are much wider than $\sim 1$~arcmin.

To map the bulk velocities and velocity dispersions, we employ forward model fitting for pre-selected sky regions (as opposed to ``backward'' image deconvolution), adopting a method first used by \cite{markevitch96a} and \cite{markevitch96b} to derive cluster temperature profiles and maps using {\it ASCA} data that was similarly affected by a broad PSF. We divided the Perseus core into seven sky regions (Sky~0 to Sky~6) as shown in figure~\ref{fig:region} right. Their combined outline extends beyond the combined outline of the FOVs of the three SXS observations as described in section~\ref{sec:velocity}, in order to keep the scattered flux from {\em outside}\/ that sky area into the FOV negligible, which is easily achieved given the cluster's sharply declining X-ray brightness profile.

We assume that the X-ray emission in each sky region is represented by a single-temperature, single-velocity thermal plasma model $M_{j}\;(j=0, 1,\dots, 6)$. As the X-ray emission from each region passes thorough the X-ray telescope, it is spread among the detector pixels because of the PSF as well as the slight drift of the satellite pointing direction during each observation. The spectra are collected in several detector regions shown in figure~\ref{fig:region} left for each of the 3 observations. The detector regions are selected to follow the sky regions as close as possible, but because of the $0.5$~arcmin pixel size and the pointing offsets, they are not the same for different observations. With Obs~1 covering only two sky Reg~ 5 and 6, we have a total of 12 spectra $S_i\;(i=1,\dots,12)$ for all regions and all pointings. Each of those spectra is the sum of the contributions from all sky regions $j$:
\begin{equation}
 S_i = R_i \sum_{j=0,6} P_{j\rightarrow i} M_{j},
\label{eq:modelsum}
\end{equation}
where $P_{j\rightarrow i}$ contains the relative flux contributions of the $j$-th sky region into the $i$-th detector region, and $R_i$\/ is the spectral redistribution matrix for the $i$-th detector region.

To calculate these relative flux contributions, we use external data --- Chandra ACIS images with a much better angular resolution. We combined Chandra ObsIDs 11713, 11714, 11715, 11716, 12025, 12033, 12036, 12037, 3209, 4289, 4946, 4947, 4948, 4949, 4950, 4951, 4952, 4953, 6139, 6145, and 6146, which include both ACIS-I and ACIS-S pointings. We used the standard Chandra data reduction techniques \citep[see, e.g.,][for details]{wang16}, including subtracting the blank-sky background after normalizing it at high energies, and modeling and subtracting the CCD readout artifact. The central AGN is a bright X-ray source affected by pileup in the ACIS image, and for our current purposes of modeling the ICM emission, we masked the central source and replaced it with the average brightness for the adjacent pixels. Other areas of the image are not affected by pileup. Point sources other than the AGN were left in the image; their flux is negligible compared to the ICM emission. We constructed two images, one in the broad 1.8--9~keV energy band and another containing only the 6.7~keV line flux --- to the accuracy possible with a CCD resolution. For the latter, we first extracted an image in the 6.4--6.7~keV band containing the redshifted 6.7~keV line complex (including the CCD line broadening). We then modeled the underlying continuum in this band by linear interpolation between images in the line-free intervals of 6.0--6.3~keV and 7.0--7.3~keV, with a small normalization correction to reflect the deviation of the spectrum from linear in this interval. The continuum image was then subtracted to result in a map of the 6.7 keV~line emission.

These Chandra images were divided into the sky regions, the image for each region was then multiplied by the mirror effective area and vignetting and convolved with the PSF for each of the 3 Hitomi pointings and at each energy of interest (the effective area, vignetting and the PSF depend on the photon energy). For each of these sky region images $j$, the flux that falls into each of the detector regions $i$\/ was collected. Technically, this was done using the Hitomi raytracing tool \verb+aharfgen+, which generates an ARF containing the values $P_{j\rightarrow i}$ in the expression above. These values for the energy of the 6.7~keV line (redshifted), are given in table~\ref{tab:psf} as fractions of the sum for all regions.

In addition to the ICM emission, the spectra have a contribution from the central AGN scattered into each integration region. Therefore, a similar calculation of the scattered contributions for a point source representing the AGN was done as above. Its normalization is determined separately using the Hitomi data (AGN~paper).

We can now derive the velocities and velocity dispersions for the 7 sky regions by fitting all 12 spectra simultaneously using the model $S_i$ that includes the ICM and AGN components. This can be done using two different technical approaches, both of which we used in this work and described them in the following section.

\subsection{The ARF method}
\label{sec:arfs}

\begin{figure*}
 \begin{center}
  \includegraphics[width=15cm]{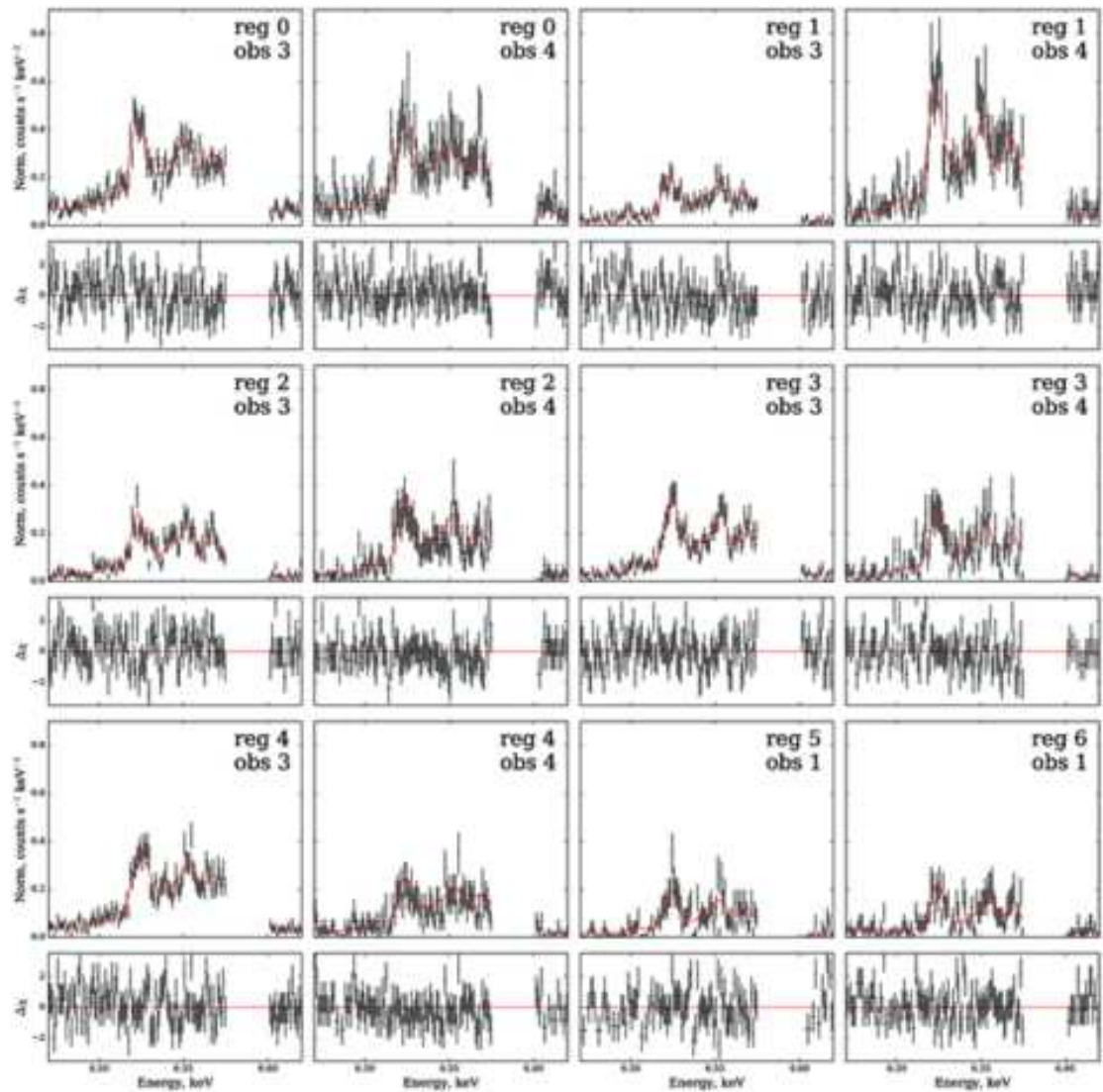}
 \end{center}
 \caption{Fits and residuals for the PSF-corrected velocity mapping.}
 \label{fig:12spec_psfcor}
\end{figure*}

In our main approach, whose results are described in section~\ref{sec:velocity}, the PSF effects are taken into account in {\small XSPEC} fitting by using the cross-region ARFs calculated as described above. This approach is general enough to allow fitting of various quantities such as temperatures and metallicities in addition to velocities. It also allows us to use different lines for velocity fitting --- e.g., excluding the resonant ({\it w}) line and using only the remaining lines of the 6.76~keV complex unaffected by resonant scattering. Because the ARF values are applied by {\small XSPEC} to the APEC normalization (as opposed to model flux), the ARF should contain values calculated using an image of the projected emission measure rather than the X-ray brightness \citep{markevitch96b}. Our broad-band Chandra image is an adequate approximation for this purpose. Furthermore, while the Chandra image contains information on the relative normalizations between various regions, given the calibration uncertainties, we let the overall model normalizations be free parameters for each spectrum. Thus we use external information only for the regions' relative contributions into each spectrum. Fitting was done in 2 steps --- first, temperatures were fit in a broad energy band excluding the 6.7~keV complex, then those temperatures were fixed, while the abundances and velocities were fit using the 6.7~keV complex (6.4--6.7~keV band excluding the {\it w} line). The best fit models and residuals for the velocity-fitting step of the above procedure are shown in figure~\ref{fig:12spec_psfcor}.

To give a clearer idea of the procedure for joint fitting of 12 spectra with 8 model components (7 plasma models and an AGN model), we show below a part of the {\small XSPEC} command file used in the velocity fitting. Note that in the current {\small XSPEC} implementation, the spectral redistribution matrix (the 'response' commands below) is specified for each of the sky region contributions, even though it is the same file for all the components within the same spectrum and could be applied after summing the model components, as shown in eq.\ (\ref{eq:modelsum}) --- this may change in the future.

\begin{lstlisting}[basicstyle=\ttfamily\scriptsize, frame=single]
 # Spectrum 1 (observation 3, approximating sky region 0):
 data 1 reg0obs3.pi

 # Contribution into this spectrum from the central point source:
 response 1:1 reg0obs3.rmf
 arf 1:1 AGN_reg0obs3.arf

 # Contribution into this spectrum from sky region 0:
 response 2:1 reg0obs3.rmf
 arf 2:1 sky0_to_reg0obs3.arf

 # Contribution into this spectrum from sky region 1:
 response 3:1 reg0obs3.rmf
 arf 3:1 sky1_to_reg0obs3.arf
 ... 
 response 8:1 reg0obs3.rmf
 arf 8:1 sky6_to_reg0obs3.arf
 
 # Spectrum 2 (observation 4, approximating sky region 0):
 data 2 reg0obs4.pi
 ...
 
 # Spectrum 12 (observation 1, approximating sky region 6):
 data 12 reg6obs1.pi
 response 1:12 reg6obs1.rmf
 arf 1:12 AGN_reg6obs1.arf
 arf 2:12 sky0_to_reg6obs1.arf
 ...
 arf 8:12 sky6_to_reg6obs1.arf

 # Spectral models: component 1 for AGN, components 2-8 for sky regions 0-6:
 model 1:agn TBabs(pegpwrlw+zgauss+zgauss)
 model 2:plasma0 TBabs*bapec
 model 3:plasma1 TBabs*bapec
 ...
 model 8:plasma6 TBabs*bapec
\end{lstlisting}

\begin{figure*}
 \begin{center}
  \includegraphics[width=15cm]{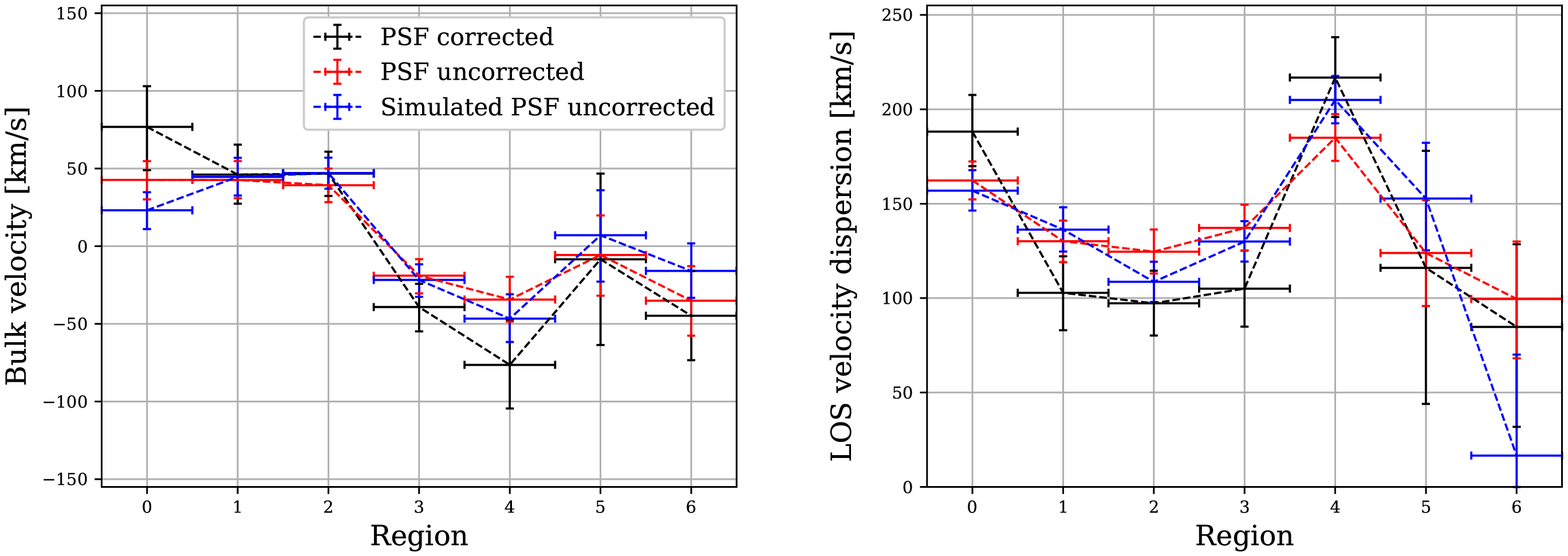}
 \end{center}
 \caption{Best-fit velocity and dispersion values for the sky regions. Black and red crosses show the PSF corrected and uncorrected fits to the real spectra, while blue crosses show the PSF-uncorrected fits to the simulated spectra, using the PSF-corrected values as input for the simulation. The agreement between blue and red crosses shows that the fitting method has found a self-consistent solution.}
 \label{fig:forward}
\end{figure*}

For a check of the results, we used the best-fit PSF-corrected values of the temperatures, abundances, velocities and dispersions, and applied the PSF blending (table~\ref{tab:psf}) and detector response to generate simulated spectra for 12 detector regions. We then fitted the simulated spectra for individual regions (without the PSF correction, but simultaneously fitting the spectra for the same sky region from different pointings). We reproduced the fits for the real spectra within the statistical errors, as shown in figure~\ref{fig:forward}.

\subsection{Simplified velocity analysis using the {\it w}\/ line}
\label{sec:maxim}

While the shape of the brightest ({\it w}) line of the He$\alpha$ triplet can be significantly affected by resonant scattering in the dense cluster core (which is indeed observed, see RS~paper), the line centroid should be less sensitive to scattering than its width. Thus, the {\it w} line can offer a useful test of the bulk velocity results derived above using the other lines of the triplet. Its width should also give an upper limit on turbulent broadening. This may be accomplished using the above ARF method, limiting the last step (fitting the velocities) to the narrow interval including only the {\it w} line. However, if we choose to fit only the {\it w} line, we can use a simpler and faster fitting approach, which is also less model-dependent, since it removes (to a good approximation) the effects of the dependence of the line flux ratios in the He$\alpha$ complex on the gas temperature.

\begin{figure*}
 \begin{center}
  \includegraphics[width=8.5cm]{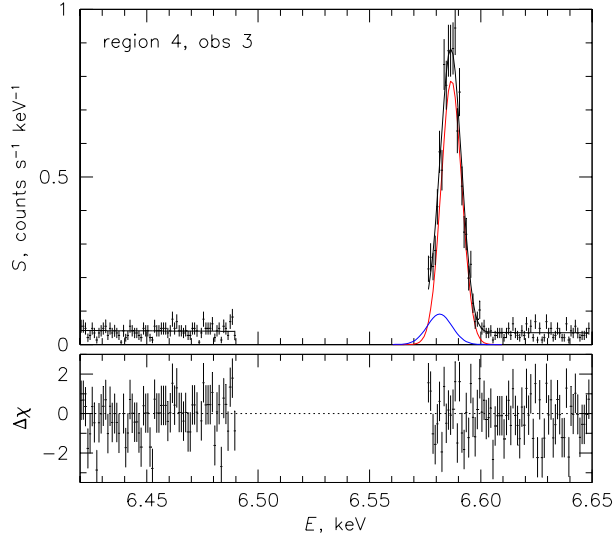}
 \end{center}  
 \vspace{15mm}
 \caption{A fit to the resonance line only, using a model consisting of a power law plus two Gaussians representing the resonance line (red curve) and a combination of its nearest satellites (blue curve). The other lines of the He$\alpha$ triplet are excluded from the fit. One spectrum is shown for illustration.}
 \label{fig:2gausreg43}
\end{figure*}

To model the {\it w} line and the underlying continuum, we fit a Gaussian plus a power law in the eneregy intervals 6.42--6.49~keV and 6.575--6.65~keV (observer frame), see figure~\ref{fig:2gausreg43}. The nearest bright component of the line complex, the {\it x} line, is 18~eV away from the {\it w} line (6.7004~keV rest frame) and is excluded using the above interval. However, there is a large number of faint satellites within $\Delta E=10$~eV of the {\it w} component, which cumulatively account for 10--15\% of the {\it w} flux (for a $T=4$~keV plasma). If not included in the model, they would bias the {\it w} line position and width. We found that these satellites can be adequately modeled by adding one Gaussian component at $E=6.695$~keV (rest frame) with an intrinsic width $\sigma=3.7$~eV and an intensity 0.138 times that of the {\it w} line. To further simplify the model, we add in quadrature the typical expected velocity dispersion of 160~km~s$^{-1}$ to this component, which gives a total width of $\sigma=5.2$~eV. While the satellite line fluxes depend on plasma temperature, and the velocty broadening is of course different in different spectra, this simplification proves to be adequate. Fitting simulated APEC spectra for a relevant range of plasma temperatures ($T=3-5$~keV) and velocity broadening ($\sigma_\mathrm{v}=100-200$~km~s$^{-1}$) using the above energy interval and a model consisting of a power law with a slope fixed at 5.0 (the local slope of the thermal spectrum for $T=4$~keV), the {\it w} line represented by a Gaussian with free redshift and width, and the satellite Gaussian with the width and relative flux fixed as above and the same redshift, we were able to recover the redshift to within 15~km~s$^{-1}$ and the line width to within 10~km~s$^{-1}$. This redshift error is acceptable given the other uncertainties, e.g., $\sim$10~km~s$^{-1}$ systematic uncertainty due to the difference between the measured \citep[e.g.][]{beiersdorfer93} and theoretical (Atomic~paper) {\it w} line energies of up to $\sim$0.3~eV. A fit to one of the spectra is shown in figure~\ref{fig:2gausreg43}, where red shows the {\it w} component and blue the satellite component. Freeing the slope of the power law does not affect the best-fit line parameters, because with our choice of the energy intervals, the continuum fit straddles the line. We also verified that fits to the real spectra using this model or full APEC in the same energy interval agree within the above errors.

We model each of the 12 spectra with a sum of 6 two-Gaussian models (one for each sky region) constructed as above. Redshifts and velocity dispersions for each sky region are tied between the 12 spectra, and the relative normalizations of the 6 main Gaussians within each spectrum are fixed to the PSF-scattered fractions given in table~\ref{tab:psf}. Here we use the fractions computed using the Chandra image of the 6.7~keV line emission (see above), which are directly applicable to our Gaussian line normalizations. Thus, instead of using 6 ARFs for each spectrum to represent the PSF contributions from each of the 6 regions, as is done in our main method (section~\ref{sec:arfs}), in this method we account for the PSF mixing within the model for each spectrum. We use only one ARF and RMF for each spectrum (we used an ARF generated for a point source in the middle of each region, but it does not matter). For reasons related to XSPEC technical implementation, this fitting method is much faster --- provided the approximations used in it are acceptable. As in the ARF method (section~\ref{sec:arfs}), we allow the overall model normalization for each spectrum to be a free parameter (even though the normalizations for each sky region can be computed from the Chandra image) to account for calibration uncertainties. The power law component for each spectrum, which represents the sum of the thermal continuum and the AGN contribution, was allowed to be a free parameter, because we are interested in the line components only and must model the underlying continuum well. It is also  possible to use APEC as a model for the {\it w} line, using the same relative model normalization scheme (though care should be taken to apply the PSF mixing fractions to {\it line fluxes} rather than the APEC normalizations), but it is much slower.

A subset of {\small XSPEC} commands for this method and a printout of the model for one of 12 spectra are given below to provide a clearer view of the procedure.

\begin{lstlisting}[basicstyle=\ttfamily\scriptsize, frame=single]

# Spectrum 1 (observation 3, approximating region 0):
data 1:1 reg0_obs3_HP_gr1.pi
response  1:1 reg0_obs3_HP_l.rmf
arf 1:1 reg0_obs3_HP_ps1890.arf

# Spectrum 2 (observation 4, approximating region 0):
data 1:2 reg0_obs4_HP_gr1.pi
response  2:2 reg0_obs4_HP_l.rmf
arf 2:2 reg0_obs4_HP_ps1890.arf
...
# Spectrum 12 (observation 1, approximating region 6)
data 1:12 reg6_obs1_HP_gr1.pi
response  12:12 reg6_obs1_HP_l.rmf
arf 12:12 reg6_obs1_HP_ps1890.arf

# Model for spectrum 1: a pair of Gaussians (a w line and the sum of the
# nearby satellites  with a normalization 0.138*w) for each of the 7 sky
# regions, plus a power law:
model 1:reg03 zgauss + zgauss + zgauss + zgauss + zgauss + zgauss + zgauss
            + zgauss + zgauss + zgauss + zgauss + zgauss + zgauss + zgauss
            + powerlaw

# Printout of the model for one spectrum (reg03), showing  parameter
# dependencies. The normalization of the `diagonal' (i=j) Gaussian 
# (parameter 4 for this spectrum) is free, while normalizations of the
# w components from other sky regions are tied to it via the relative 
# PSF contributions (for simplicity we use 0 for the fractions <5%):
   1    1   zgauss     LineE      keV      6.70040      frozen
   2    1   zgauss     Sigma      keV      5.41827E-03  +/-  4.01105E-04  
   3    1   zgauss     Redshift            1.76995E-02  +/-  5.97538E-05  
   4    1   zgauss     norm                5.19561E-05  +/-  1.72050E-06  
   5    2   zgauss     LineE      keV      6.69500      frozen
   6    2   zgauss     Sigma      keV      5.20000E-03  frozen
   7    2   zgauss     Redshift            1.76995E-02  = reg03:p3
   8    2   zgauss     norm                7.16994E-06  = reg03:p4*0.138
   9    3   zgauss     LineE      keV      6.70040      frozen
  10    3   zgauss     Sigma      keV      3.39663E-03  +/-  2.76307E-04  
  11    3   zgauss     Redshift            1.74271E-02  +/-  4.28597E-05  
  12    3   zgauss     norm                8.41689E-06  = reg03:p4*0.162
  13    4   zgauss     LineE      keV      6.69500      frozen
  14    4   zgauss     Sigma      keV      5.20000E-03  frozen
  15    4   zgauss     Redshift            1.74271E-02  = reg03:p11
  16    4   zgauss     norm                1.16153E-06  = reg03:p12*0.138
  17    5   zgauss     LineE      keV      6.70040      frozen
  18    5   zgauss     Sigma      keV      3.48743E-03  +/-  2.36872E-04  
  19    5   zgauss     Redshift            1.75454E-02  +/-  3.76188E-05  
  20    5   zgauss     norm                1.15343E-05  = reg03:p4*0.222
  21    6   zgauss     LineE      keV      6.69500      frozen
  22    6   zgauss     Sigma      keV      5.20000E-03  frozen
  23    6   zgauss     Redshift            1.75454E-02  = reg03:p19
  24    6   zgauss     norm                1.59173E-06  = reg03:p20*0.138
  25    7   zgauss     LineE      keV      6.70040      frozen
  26    7   zgauss     Sigma      keV      4.65027E-03  +/-  2.67985E-04  
  27    7   zgauss     Redshift            1.71175E-02  +/-  4.26310E-05  
  28    7   zgauss     norm                6.18277E-06  = reg03:p4*0.119
  29    8   zgauss     LineE      keV      6.69500      frozen
  30    8   zgauss     Sigma      keV      5.20000E-03  frozen
  31    8   zgauss     Redshift            1.71175E-02  = reg03:p27
  32    8   zgauss     norm                8.53223E-07  = reg03:p28*0.138
  33    9   zgauss     LineE      keV      6.70040      frozen
  34    9   zgauss     Sigma      keV      3.95278E-03  +/-  2.31365E-04  
  35    9   zgauss     Redshift            1.71612E-02  +/-  3.71691E-05  
  36    9   zgauss     norm                5.09170E-06  = reg03:p4*0.098
  37   10   zgauss     LineE      keV      6.69500      frozen
  38   10   zgauss     Sigma      keV      5.20000E-03  frozen
  39   10   zgauss     Redshift            1.71612E-02  = reg03:p35
  40   10   zgauss     norm                7.02654E-07  = reg03:p36*0.138
  41   11   zgauss     LineE      keV      6.70040      frozen
  42   11   zgauss     Sigma      keV      4.61808E-03  +/-  6.04779E-04  
  43   11   zgauss     Redshift            1.73606E-02  +/-  9.80638E-05  
  44   11   zgauss     norm                0.0          = reg03:p4*0.
  45   12   zgauss     LineE      keV      6.69500      frozen
  46   12   zgauss     Sigma      keV      5.20000E-03  frozen
  47   12   zgauss     Redshift            1.73606E-02  = reg03:p43
  48   12   zgauss     norm                0.0          = reg03:p44*0.138
  49   13   zgauss     LineE      keV      6.70040      frozen
  50   13   zgauss     Sigma      keV      3.92239E-03  +/-  4.30404E-04  
  51   13   zgauss     Redshift            1.70599E-02  +/-  6.30424E-05  
  52   13   zgauss     norm                0.0          = reg03:p4*0.
  53   14   zgauss     LineE      keV      6.69500      frozen
  54   14   zgauss     Sigma      keV      5.20000E-03  frozen
  55   14   zgauss     Redshift            1.70599E-02  = reg03:p51
  56   14   zgauss     norm                0.0          = reg03:p52*0.138
  57   15   powerlaw   PhoIndex            5.00000      frozen
  58   15   powerlaw   norm                7.22371      +/-  0.208572     

# In models for other spectra, redshifts and widths of the lines are
# tied to the values for the respective sky region in this model. 
# A total of 38 parameters are being fit.
\end{lstlisting}

\begin{table*}
  \tbl{Best-fit bulk velocity and LOS velocity dispersion values. Values of
    $v$ and $\sigma_\mathrm{v}$ are km\,s$^{-1}$.}{%
  \begin{tabular}{crrrrrrrrrrr}
   \hline
   & \multicolumn{5}{c}{{\it w}\/ line excluded, ARF method} & 
   & \multicolumn{5}{c}{Fit {\it w}\/ line only, model mixing method}\\
   \cline{2-6} \cline{8-12}
   & \multicolumn{2}{c}{PSF uncorrected} && \multicolumn{2}{c}{PSF corrected} & & \multicolumn{2}{c}{PSF uncorrected} && \multicolumn{2}{c}{PSF corrected}\\
  region & $v_\mathrm{bulk}$ & $\sigma_\mathrm{v}$ & & $v_\mathrm{bulk}$ & $\sigma_\mathrm{v}$ & & $v_\mathrm{bulk}$ & $\sigma_\mathrm{v}$ & & $v_\mathrm{bulk}$ & $\sigma_\mathrm{v}$ \\
  \hline
  0  & $43_{-13}^{+12}$  & $163_{-10}^{+10}$ &~~~~~~& $75_{-28}^{+26}$  & $189_{-18}^{+19}$ &~~~~~~~~~&$50_{-8}^{+8}$&$194_{-8}^{+9}$&~~~~~~&$98_{-18}^{+20}$  & $228_{-20}^{+21}$\\
  1  & $42_{-12}^{+12}$  & $131_{-11}^{+11}$ &      & $46_{-19}^{+19}$  & $103_{-20}^{+19}$ & & $33_{-8}^{+8}$  &   $163_{-9}^{+9}$   & & $16_{-13}^{+13}$   & $127_{-16}^{+17}$ \\
  2  & $39_{-11}^{+11}$  & $126_{-12}^{+12}$ &      & $47_{-14}^{+14}$  & $98_{-17}^{+17}$  & & $39_{-7}^{+8}$    & $158_{-9}^{+9}$   & & $52_{-12}^{+11}$   & $132_{-14}^{+15}$ \\
  3  & $-19_{-11}^{+11}$ & $138_{-12}^{+12}$ &      & $-39_{-16}^{+15}$ & $106_{-20}^{+20}$ & & $-43_{-8}^{+8}$   & $193_{-8}^{+9}$   & & $-76_{-13}^{+13}$  & $191_{-16}^{+16}$ \\
  4  & $-35_{-14}^{+15}$ & $186_{-12}^{+12}$ &      & $-77_{-28}^{+29}$ & $218_{-21}^{+21}$ & & $-30_{-7}^{+7}$   & $175_{-7}^{+8}$   & & $-63_{-11}^{+11}$  & $156_{-14}^{+14}$\\
  5  & $-6_{-26}^{+25}$  & $125_{-28}^{+28}$ &      & $-9_{-56}^{+55}$  & $117_{-73}^{+62}$ & & $-9_{-17}^{+17}$  & $175_{-18}^{+20}$ & & $-3_{-29}^{+30}$  & $189_{-35}^{+37}$ \\
  6  & $-35_{-22}^{+22}$ & $99_{-32}^{+31}$  &      & $-45_{-29}^{+29}$ & $84_{-54}^{+44}$  & & $-78_{-15}^{+15}$ & $164_{-16}^{+17}$ & & $-93_{-19}^{+19}$ & $154_{-20}^{+21}$ \\
     \hline
  \end{tabular}}\label{tab:velocity_modelmix}
\end{table*}

\begin{figure*}
 \begin{center}
  \includegraphics[width=15cm]{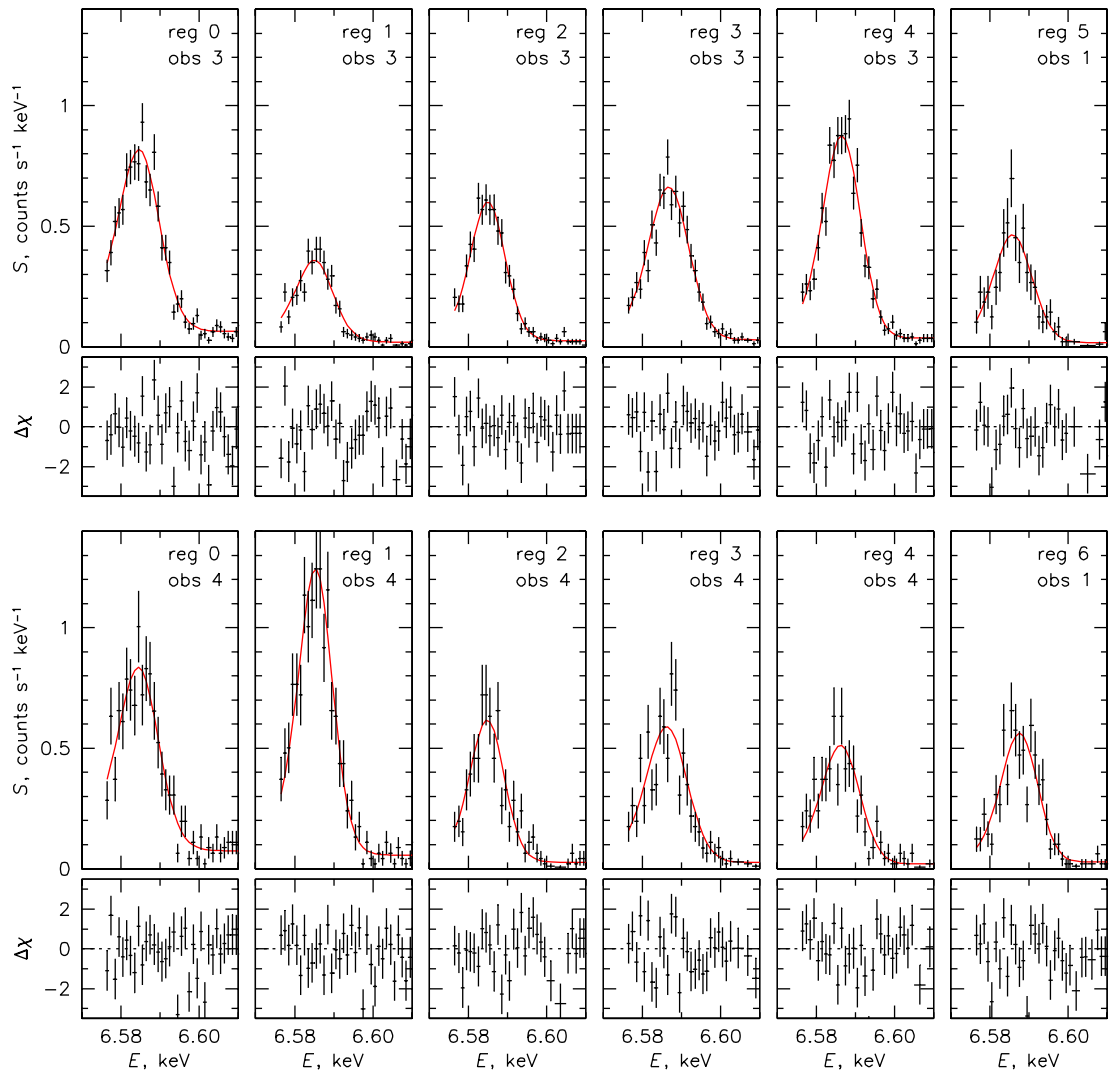}
 \end{center}
 \vspace{30mm}
 \caption{Fits and residuals for the joint fit of all spectra using the {\it w}-line method (figure~\ref{fig:2gausreg43}). The continuum energy interval to the left of the line complex (see figure~\ref{fig:2gausreg43}) is not not shown.}
 \label{fig:12spec}
\end{figure*}

The resulting LOS velocities (in the same reference frame as above) and velocity dispersions are given in table~\ref{tab:velocity_modelmix}. It gives both the individual, PSF-uncorrected fits for each sky region (fitting together either two or one spectra for each region, as above) and the joint PSF-corrected fit to all spectra. The joint fit is good, with C statistic of 1390 for 1484 d.o.f. Residuals for individual spectra are shown in figure~\ref{fig:12spec}.

We note that these velocities and dispersions are derived using both a different fitting method and the independent data excluded from our main fit. It thus provides a good check of that fit. The PSF-corrected velocities from both methods are in good statistical agreement and show the same large-scale velocity gradient. The velocity dispersions from the {\it w} line method show approximately similar spatial pattern, but most values are higher (though they are statistically inconsistent only in Reg~3). Higher widths for the {\it w} line are expected in the presence of resonant scattering. We also fit the {\it w} line in the same energy interval using the ARF method and obtained results very close to those from the simplified method.

\subsection{Velocity analysis using He$\beta$ lines}
\label{sec:heb}

Fe He$\beta$ lines are optically thin and thus provide another consistency check of our main result in section~\ref{sec:velocity}. We focused our comparison only on the PSF uncorrected $\sigma_\mathrm{v}$ values, because the statistics of the He$\beta$ lines is not as good as that of the He$\alpha$ complex and the gain is not well calibrated compared to the He$\alpha$ complex.

We fit the spectra in the energy range of 7.7--7.8~keV or 7.6--7.9~keV (observer frame) using \verb+bapec+, with all parameters allowed to vary. The narrower energy range includes only the He$\beta$ lines and the results are not significantly affected by other lines. The results obtained using the wider energy range are not as clean as the former ones, because the energy range also covers the Ni He$\alpha$ line, but the errorbars are small because of the higher statistics for the continuum determination. As in above PSF-uncorrected fits, we fit spectra from different sky regions independently, while fitting simultaneously the spectra for the same sky region from different pointings.

\begin{table*}
 \tbl{Best-fit LOS velocity dispersion ($\sigma_\mathrm{v}$) in the unit of km~s$^{-1}$.}{%
 \begin{tabular}{crrr}
  \hline
  & He$\alpha$, {\it w}\/ excluded & He$\beta$, 7.7--7.8~keV & He$\beta$, 7.6--7.9~keV\\
  \hline
  0  & $163_{-10}^{+10}$ &  $145_{-32}^{+36}$ & $160_{-29}^{+31}$\\
  1  & $131_{-11}^{+11}$ & $112_{-112}^{+31}$ & $109_{-26}^{+25}$\\
  2  & $126_{-12}^{+12}$ &   $95_{-83}^{+62}$ & $154_{-28}^{+30}$\\
  3  & $138_{-12}^{+12}$ &  $152_{-32}^{+31}$ & $138_{-30}^{+31}$\\
  4  & $186_{-12}^{+12}$ &  $196_{-25}^{+27}$ & $184_{-23}^{+24}$\\
  5  & $125_{-28}^{+28}$ &  $42_{-42}^{+128}$ & $151_{-77}^{+96}$\\
  6  & $99_{-32}^{+31}$  &   $96_{-96}^{+95}$ & $171_{-79}^{+70}$\\
  \hline
 \end{tabular}}\label{tab:heb}
\end{table*}

The resulting best-fit $\sigma_\mathrm{v}$ are shown in table~\ref{tab:heb}. The PSF-uncorrected $\sigma_\mathrm{v}$ values obtained from the optically thin lines (He$\alpha$ {\it x}+{\it y}+{\it z} or He$\beta$) are consistent with each other.

\end{document}